\documentclass[structabstract]{aa}

\usepackage{txfonts}
\usepackage{graphicx}
\usepackage{natbib}
\bibpunct{(}{)}{;}{a}{}{,}

\begin{document}

  \title{A close look at the Centaurus A group of galaxies \\ 
III. Recent star formation histories of late-type dwarfs around M83}
  \author{D. Crnojevi\'{c}\inst{1}\fnmsep\thanks{Corresponding \email{denija@ari.uni-heidelberg.de}. \newline Member of IMPRS (International Max Planck Research School) for Astronomy \& Cosmic Physics at the University of Heidelberg and of the Heidelberg Graduate School for Fundamental Physics.} \and E. K. Grebel\inst{1} \and A. A. Cole\inst{2}
}

 \institute{Astronomisches Rechen-Institut, Zentrum f\"{u}r Astronomie der
   Universit\"{a}t Heidelberg, M\"{o}nchhofstrasse 12-14, 69120 Heidelberg, Germany
 \and
School of Mathematics \& Physics, University of Tasmania, Private Bag 37 Hobart, 7001 Tasmania, Australia}

  \date{Received 27 July 2010 / Accepted 17 February 2011}

  \abstract
    {}{We study the resolved stellar populations of dwarf galaxies in the nearby Centaurus A/M83 group of galaxies. Our goal is to characterize their evolutionary history and to investigate eventual similarities or differences with the dwarf population in other group environments.}{This work presents the analysis of five late-type (irregular) dwarfs found in the vicinity of the giant spiral M83. Using archival HST/ACS data, we perform synthetic color-magnitude diagram modeling to derive the star formation histories of these late-type dwarfs.}{The target objects show heterogeneous star formation histories, with average star formation rates of 0.08 to $0.70\times10^{-2}$M$_\odot$ yr$^{-1}$. Some of them present prolonged, global bursts of star formation ($\sim300-500$ Myr). The studied galaxies are all metal-poor ([Fe/H]$\sim-1.4$). 
We further investigate the spatial extent of different stellar populations, finding that the young stars show a clumpy distribution, as opposed to the smooth, broad extent of the old ones. The actively star forming regions have sizes of $\sim100$ pc and lifetimes of $\gtrsim100$ Myr, thus suggesting a stochastic star formation mode for the target dwarf irregular galaxies. The  galaxies formed $\sim20\%$ to $70\%$ of their stars more than $\sim7$ Gyr ago.}{The studied dwarfs have average star formation rates slightly higher than their analogues in the Local Group, but comparable to those in the M81 group. Our preliminary sample indicates that the neutral gas content of the target dwarfs does seem to be affected by the group environment: galaxies within a denser region have a much lower $M_{HI}/<SFR>$ than the isolated ones, meaning that they will exhaust their gas reservoir more quickly. }

\titlerunning{A close look at the Centaurus A group of galaxies III. Late-type dwarfs around M83}

  \maketitle

%________________________________________________________________

\section{Introduction}

Since the first detailed studies of the smaller counterparts of the well-studied giant galaxies, dwarf galaxies have revealed an unexpected amount of intriguing properties. They are by far the most numerous galaxy population in our Universe, with many new faint galaxies being recently discovered in our Local Group \citep[see, e.g.,][]{belokurov06, zucker06b, zucker06a, zucker07, belokurov10}. Dwarf galaxies are possibly the baryonic counterpart of dark matter building blocks, predicted by $\Lambda$CDM cosmology. From spectroscopic studies of dwarf satellites around our Milky Way, we also know that they appear to be strongly dark matter dominated \citep[e.g., ][]{gilmore07}. None of the Local Group dwarfs studied in detail so far lacks an ancient population, although the fraction of ancient stars varies strongly from galaxy to galaxy \citep[e.g.,][]{grebel04, orban08}.

Many efforts have been made to analyze the physical properties of the dwarf companions of the Milky Way and of M31, which are ideal targets due to their proximity. We know for example that different morphological types exhibit very different characteristics (see \citealt{grebel01} for a review). Early-type dwarfs consist of predominantly old stellar populations ($>10$ Gyr), and are normally depleted in HI gas. However, there are also examples (Carina, Fornax, LeoI, LeoII) where intermediate-age populations provide a good fraction of the stellar mass. On the other side, late-type dwarfs (irregulars) always show the presence of young stars ($<1$ Gyr) on top of the older populations, and they are normally gas-rich and more luminous. Additionally, dwarf irregulars are usually found at greater distances from the giant dominant galaxy, whereas early-type dwarfs are located mostly within the inner $\sim300$ kpc of the group center \citep[e.g., ][]{einasto74, dressler80, kara02}. All dwarf galaxies tend to be metal-poor, with average [Fe/H] values of $\lesssim-1.0$ dex. In particular, both types follow a metallicity-luminosity relation, such that more luminous objects have also a higher mean metal content, but late-type dwarfs appear to be more metal-poor at a fixed luminosity \citep[e.g.,][]{grebel03}. Finally, the so called transition-type dwarfs show stellar population characteristics similar to those of early-type dwarfs, but also contain some neutral gas, just as late-type dwarfs, with the difference that they do not currently form stars.

One of the perhaps most astonishing properties of such small objects is that there are no two dwarfs that are alike, as they exhibit a wide range of diverse evolutionary histories \citep{grebel97, tolstoy09}. Deep color-magnitude diagrams (CMDs) permit us to derive the detailed star formation histories (SFHs) for dwarf galaxies in the Local Group. In particular, the technique of synthetic CMD modeling was pioneered some twenty years ago, and since then it has been further developed and improved by many different groups \citep[e.g., ][]{tosi91, aparicio96, dolphin02, cole07}. We are thus able to put strong constraints on the star formation epochs and efficiency for dwarf galaxies, provided that the photometry reaches the main sequence turnoff of the old populations. These kinds of analyses were preferentially performed on early-type dwarfs, since they require long exposure times and thus are easier to apply for closer objects. As for late-type dwarfs, generally found in the outskirts of the Local Group, their CMD-modeling census has not yet been completed. For example, detailed SFHs have already been determined from deep CMDs for the Magellanic Clouds \citep[e.g.,][]{smecker02, harriszar04, harriszar09, noel09, sabbi09}, IC1613 \citep{skillman03}, Leo A \citep{cole07}, IC10 \citep{cole09}. From these and other works we learn that late-type dwarfs tend to form stars slowly over long periods, separated by short quiescent phases \citep[e.g.,][and references therein]{cignoni10}. Spatial studies of the star formation in these small galaxies, both within and beyond the Local Group, have also been performed \citep[e.g.,][]{paynegap72, issers84, dohm97, vandyk98, dohm02, weisz08, glatt10}, leading to the conclusion that their star formation appears to be stochastic. This picture gains further support from abundance studies, which indicate that these dwarfs are not well-mixed \citep[e.g., ][]{kniazev05, glatt08a, koch08b}.

Beyond the Local Group, the integrated properties of dwarf galaxies in nearby groups have also been studied extensively \citep[e.g.,][]{trentham02, makarova02, kara04, kara05, makarova05, sharina08, bouchard08, koleva09}. However, their stellar populations can be studied in detail only with advanced space instrumentation (Hubble Space Telescope), and even in those cases, only the brightest stars are resolvable out to a distance of $\sim17$ Mpc \citep[e.g.][]{caldwell06, durrell07, williams07}. The first attempts of deriving SFHs via synthetic modeling beyond the Local Group were already made more than a decade ago \citep[e.g., ][]{tosi89, tosi91, dohm97, aloisi99, schulte99, aloisi01, makarova02}, and recently there has been an increasing effort in this kind of studies \citep[e.g.,][and references therein]{weisz08, mcquinn09, tolstoy09}, but we are still very far from having a complete census. Even though we can only obtain photometric depths that are comparable to the ones we used to reach for objects in the Local Group about one decade ago, the results can still shed light on how dwarf galaxies evolve in different neighbourhoods. Are the SFHs influenced by the surrounding environment? Until now, there is no evidence for large differences among dwarf irregulars within diverse galaxy groups (see, for example, the comparison between the Local Group and the M81 group in \citealt{weisz08}), which could lead us to the conclusion that internal processes are the ones governing the star formation in these objects. On the other hand, the dwarf galaxy types do seem to depend on environment. For example, very loose groups such as the Sculptor group and Canes Venatici Cloud host primarily late-type dwarfs \citep[see ][]{kara03_can, kara03_sc}, whereas more evolved, higher density groups like the Local Group, the M81 and CenA groups have a sizeable fraction of early-type dwarfs \citep{kara02_m81, kara02}.

The Centaurus A/M83 group has a mean distance from us of $\sim4$ Mpc, and thus its members are still resolvable into stars. The complex is formed by two smaller subgroups, whose dominant galaxies are Centaurus A (CenA) and M83, even though it is not clear whether they are receding from or approching each other \citep[e.g., ][]{kara07}. The group contains about 50 dwarf galaxies, and with respect to the Local Group it is a denser and possibly more evolved environment (see discussion in \citealt{crnojevic10}). The dwarf population has already been studied in the past \citep{cote97, kara98, banks99, cote00, jerjen00b, jerjen00a, kara02, lee03, kara04, kara05, rejkuba06, bouchard07, grossi07, kara07, lee07, makarova08, sharina08, bouchard08, makarova09, cote09}: most of these works consider the physical properties of large samples of objects at different wavelengths. The main conclusions from these datasets are that the scaling relations (e.g., morphology-density, metallicity-luminosity) are comparable to those observed in the Local Group and in other nearby groups. The Centaurus A group environment is supposed to be efficient in the gas stripping of its dwarf members \citep{bouchard07}, but some of its late-type dwarfs contain an unexpectedly large fraction of neutral gas \citep{grossi07}. The present-day star formation efficiencies of its dwarf irregulars in general are lower in denser regions of the group \citep{bouchard08}. Despite the fact that the giant members of the group all show elevated star formation rates (SFRs) and hints for recent interaction events, the H$\alpha$ fluxes of the late-type dwarf population do not reveal any sign of enhanced activity with respect to other groups \citep{cote09}.

In the present work, we want to look more in detail at the SFRs over the late-type dwarf galaxies' lifetimes, taking advantage of the resolved stellar populations observed with the Hubble Space Telescope. For the predominantly old early-type dwarfs of the Centaurus A group, the information available from the CMD is rather limited, but we were able to photometrically estimate their metallicities. We find population gradients in some of these objects \citep{crnojevic10}, and estimate their intermediate-age population fractions (Crnojevi\'{c} et al. 2011, submitted). The late-type dwarfs contain also young stellar components, and it is thus possible to perform synthetic CMD modeling to derive their recent SFHs. We do so for 5 dwarfs that are found within and around the M83 subgroup. In addition, we look for potential environmental influences on the derived properties. A further sample of 5 objects located in the CenA surroundings will be presented in a companion paper (Crnojevi\'{c}, Grebel \& Cole 2011b, in prep.).

The present paper is organized as follows: we describe the data in \S \ref{data_sec}, and present the derived results in \S \ref{cmd_sec} (color-magnitude diagrams), \S \ref{sfh_sec} (SFHs), and \S \ref{maps_sec} (population gradients). The discussion is then carried out in \S \ref{discuss}, and our conclusion are drawn in \S \ref{conclus}.

%________________________________________________________________

\section{Data and photometry} \label{data_sec}

\begin{table*}
 \centering
\caption{Fundamental properties of the studied sample of galaxies.}
\label{infogen}
\begin{tabular}{lcccccccccc}
\hline
\hline
Galaxy&RA&DEC&$D$&$(m-M)_{0}$&$D_{M83}$&$\theta$&$A_{I}$&$M_{B}$&$M_{HI}$&$\Theta$\\
&(J2000)&(J2000)&(Mpc)&&(kpc)&($^{\circ}$)&&&$(10^6$M$_{\odot})$\\
\hline
 \object{ESO381-18}&$12\,44\,42.7$&$-35\,58\,00$&$5.32\pm0.51$&$28.63\pm0.14$&$1156\pm125$&$12.54$&$0.12$&$-12.91$&$27/29$\tablefootmark{a,b}&$-0.6$\\
 \object{ESO381-20}&$12\,46\,00.4$&$-33\,50\,17$&$5.44\pm0.37$&$28.68\pm0.14$&$1100\pm144$&$11.53$&$0.13$&$-14.44$&$157/251$\tablefootmark{a,b}&$-0.3$\\
 \object{ESO443-09, KK170}&$12\,54\,53.6$&$-28\,20\,27$&$5.97\pm0.46$&$28.88\pm0.17$&$1212\pm415$&$9.32$&$0.13$&$-11.82$&$14$\tablefootmark{b}&$-0.9$\\
 \object{IC4247, ESO444-34}&$13\,26\,44.4$&$-30\,21\,45$&$4.97\pm0.49$&$28.48\pm0.21$&$277\pm437$&$2.28$&$0.12$&$-14.07$&$34/37$\tablefootmark{a,b}&$1.5$\\
 \object{ESO444-78, UGCA365}&$13\,36\,30.8$&$-29\,14\,11$&$5.25\pm0.43$&$28.60\pm0.17$&$107\pm499$&$0.64$&$0.10$&$-13.11$&$18/21$\tablefootmark{b,c}&$2.1$\\
\hline
\end{tabular}
\tablefoot{
The columns are the following: (1): name of the galaxy; (2-3): equatorial coordinates from \citet{kara07} (J2000; units of right ascension are hours, minutes and seconds, and units of declination are degrees, arcminutes and arcseconds); (4-5): distance and distance modulus of the galaxy derived by \citet{kara07} with the tip of the red giant branch method; (6-7): deprojected and angular distance from M83; (8): Galactic foreground extinction in the $I$-band from \citet{schlegel98}; (9): absolute $B$ magnitude (converted from \citealt{bouchard08} with the distance modulus listed in column (6)); (10): HI mass obtained from different sources (\tablefoottext{a}{\citealt{banks99}, recomputed with updated values of the distance}; \tablefoottext{b}{\citealt{georgiev08}}; \tablefoottext{c}{\citealt{bouchard08}}); and (11): tidal index (i.e., degree of isolation), taken from \citet{kara07}.
}
\end{table*}

The five target galaxies of this study have been observed by the Wide Field Channel (WFC) of the Advanced Camera for Surveys (ACS) aboard the Hubble Space Telescope (HST), during the observing programmes GO-9771 and GO-10235. For each galaxy there is a 1200 seconds exposure time in the $F606W$ filter (corresponding to the broad $V$-band in the Johnson-Cousins system) and a 900 seconds exposure time in the $F814W$ filter (corresponding to the broad $I$-band). 

We perform resolved stellar photometry with the DOLPHOT package \citep{dolphin02}, using the ACS module and adopting the parameters suggested in the User's Guide\footnote{http://purcell.as.arizona.edu/dolphot/.}. The output photometry is given both in the instrumental and in the Johnson-Cousins system (converted by the program following \citealt{sirianni05}). In order to reject non-stellar detections and to have a clean final sample of stars for our CMDs, we apply quality cuts. We require the signal-to-noise to be $\geq5$, $\chi^2\leq2.5$, the sharpness to be between $-0.3$ and 0.3 to avoid too sharp (cosmic rays or detector defects) or too extended objects (like background galaxies), and finally the crowding parameters (which quantify how much brighter the star would be if isolated when measured) to be $\leq0.5$ in both filters. However, for some of the galaxies (ESO381-18 and IC4247) the standard parameters for the photometry do not give satisfying results. More specifically, the stars found in central, crowded regions are rejected by our adopted quality cuts. For these galaxies, we change the DOLPHOT photometry parameter Force1 from the value suggested in the User's Guide. This would normally force the code to retain only ``good'' objects, meaning not too faint for the point spread function determination, not too sharp or too elongated. By changing this parameter, also sharp/extended objects are classified as ``good'' objects in the output of the photometry, so that the final photometry will now retain valid objects in the crowded regions, but will also add spurious objects. To avoid the latter, we apply stricter quality cuts (using $-0.2\leq$ sharpness $\leq0.2$), and remove spurious objects by hand when this is not enough (e.g., in the cases where pointlike objects are detected in the tails of saturated foreground stars). In this way, we do not lose precious information about the bluest part of the CMD, with the only disadvantage of having a less clean diagram (i.e., the main sequence/blue helium burning sequence and the red giant branch features show a larger scatter at magnitudes $I\gtrsim24.5$, and more spurious objects are found at colors $V-I\gtrsim2$).

The DOLPHOT output includes photometric errors for each star, however, these errors do not account for systematic errors of the point spread function. Thus, extensive artificial star tests are performed for each of the studied objects, using the same photometry and quality cut parameters of the original image, in order to estimate the photometric uncertainties and to constrain the parameters derived from our SFH recovery (see Sect. \ref{sfh_sec}). We add $\sim5-10$ times the number of real stars (after quality cuts), and we distribute them evenly across the field of view and such that they cover the whole color and magnitude range of the observed stars. Moreover, the artificial stars are simulated also at magnitudes below our detection limit, in order to take into account objects that are actually fainter than what we observe, but that are detected because of an addition of noise. The photometry of the artificial stars is performed by the program for one star at a time, so that we are not artificially introducing over-crowding. At a $50\%$ completeness level, the limiting magnitude for the least crowded galaxy (ESO443-09, with a peak density of $\sim 2$ stars per arcsec$^{2}$, corresponding to $\sim167$ stars per $0.1$ kpc$^{2}$) is $\sim27.5$ mag ($\sim26.8$ mag) for the $V$-band ($I$-band). At the mentioned $I$-band magnitude, the representative $1 \sigma$ photometric error (as derived from artificial star tests) amounts to $\sim0.18$ mag in magnitude and $\sim0.23$ mag in color. For the most crowded one of our targets (IC4247, peak density of $\sim 10$ stars per arcsec$^{2}$, or $\sim1041$ stars per $0.1$ kpc$^{2}$), the $50\%$ completeness level is reached at $\sim26.9$ mag ($\sim26.2$ mag) for the $V$-band ($I$-band), while the corresponding $1 \sigma$ photometric errors for this $I$-band magnitude are $\sim0.21$ mag in magnitude and $\sim0.28$ mag in color. We show the mean photometric errors for each galaxy in the CMDs of the next Sections.

We list the general properties of the target galaxies in Table \ref{infogen}. We additionally note that the morphological de Vaucouleurs type is the same for all the galaxies and has a value of 10. Moreover, we use the tidal index parameter throughout this work, as defined by \citet{kara99}:
\begin{displaymath}
\Theta_{i} = \mathrm{max} \{\mathrm{log}(M_{k}/D^{3}_{ik})\} + C, \;\;\;	k = 1,2,...N,
\end{displaymath}
for the galaxy $i$, where $M_{k}$ and $D_{ik}$ refer to the mass and deprojected distance of any of its neighboring galaxies, respectively. This parameter thus quantifies the maximum density enhancement produced by the companions of the object in study. The value of $C$ is set for each galaxy such that $\Theta=0$ when the Keplerian orbit of the galaxy with respect to its main disturber has a period equal to a cosmic Hubble time.

\begin{figure*}
 \centering
  \includegraphics[width=17.5cm]{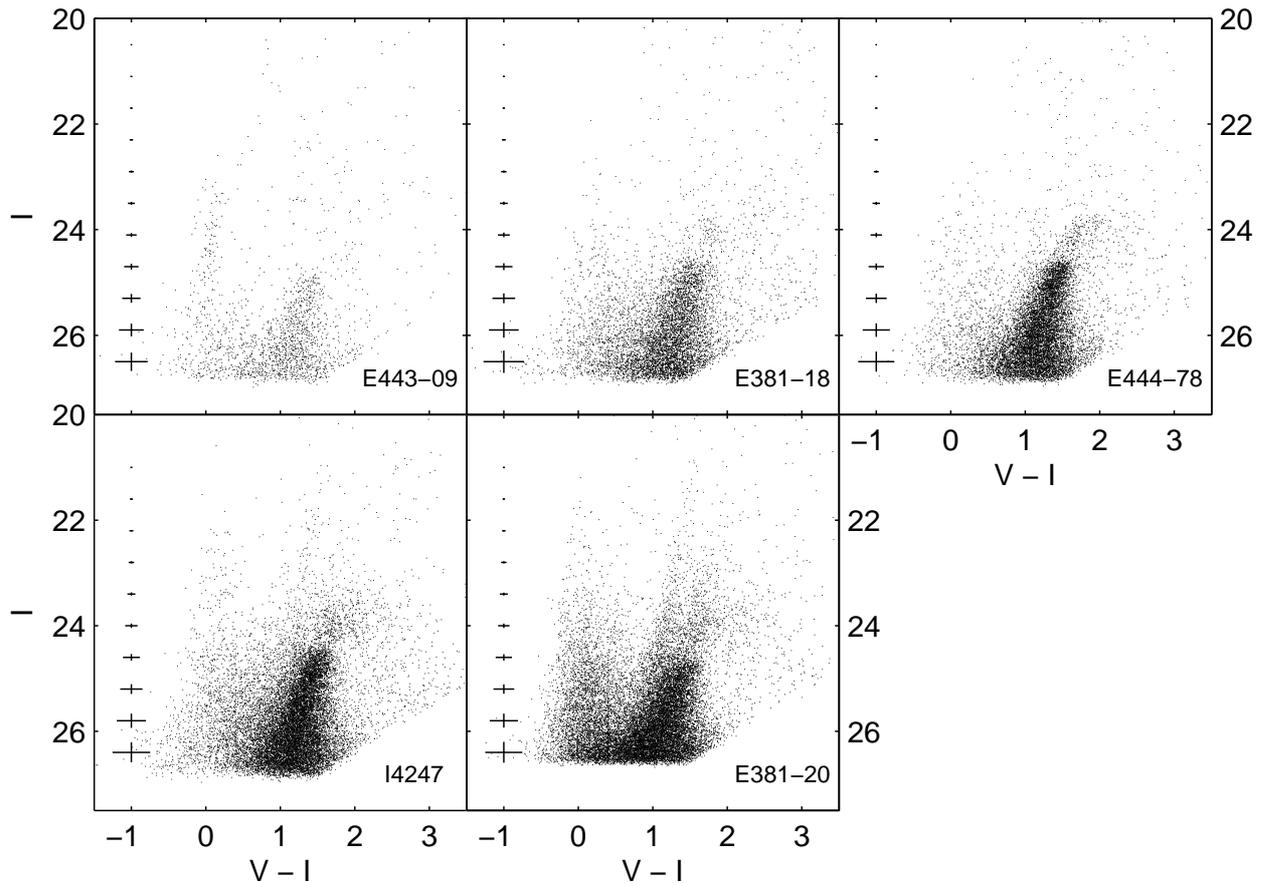}
 \caption{\footnotesize{Color-magnitude diagrams of the five late-type dwarf galaxies studied here, ordered by (increasing) absolute magnitude. The main features visible in all of the diagrams are the blue plume (main sequence and blue helium-burning stars), the upper red giant branch, the luminous asymptotic giant branch (in all but for ESO443-09) and for IC4247 and ESO381-20 also a prominent red supergiant region (see text for details). On the left side of each diagram, representative $1\sigma$ photometric errorbars (as derived from artificial star tests) are shown.}}
 \label{fore}
\end{figure*}

\subsection{Galactic foreground contamination} \label{fore_sec}

\begin{figure*}
 \centering
  \includegraphics[width=17.5cm]{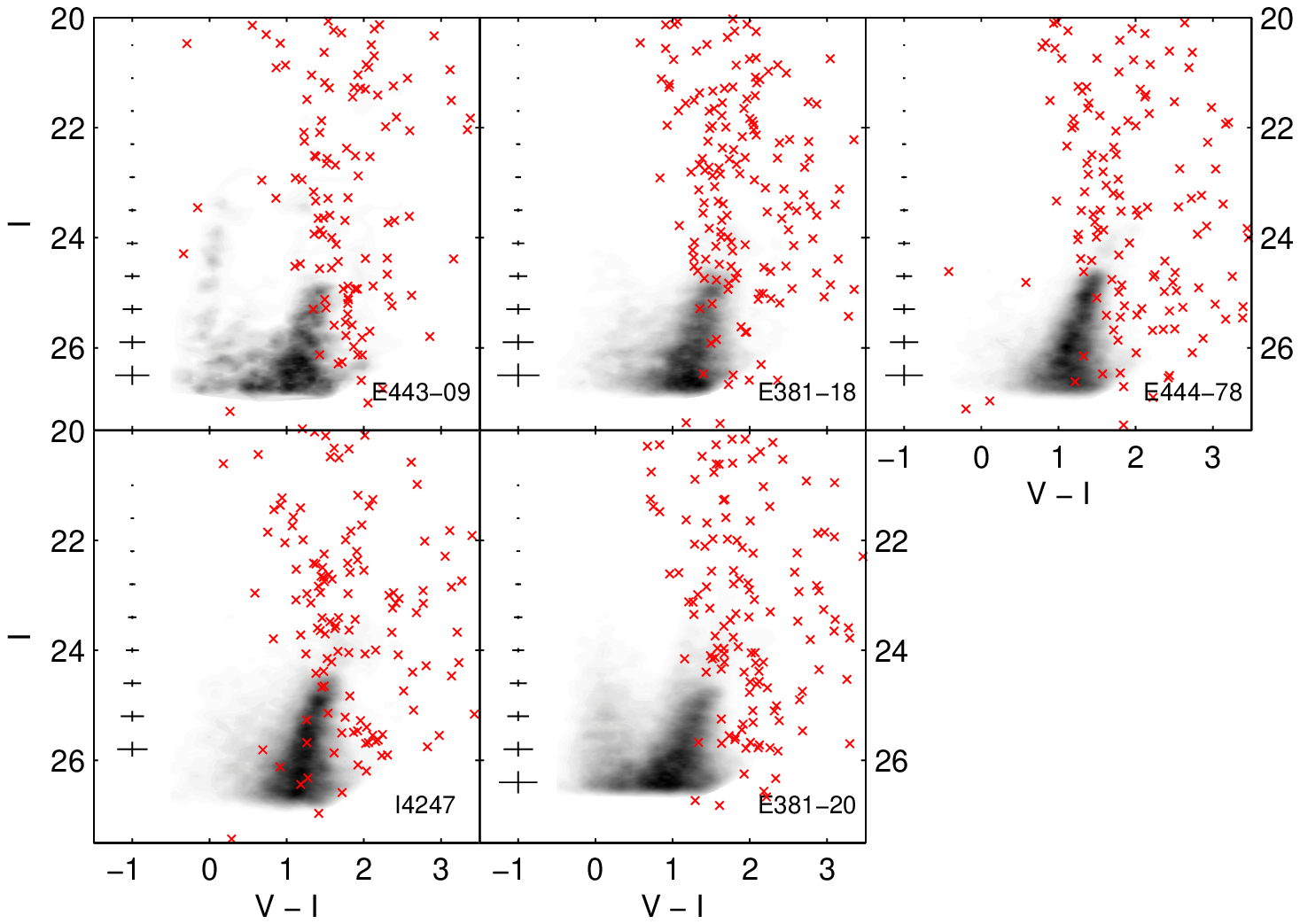}
 \caption{\footnotesize{Hess density diagrams of the five late-type dwarf galaxies studied here, ordered by (increasing) absolute magnitude. We additionally overplot to each CMD the Galactic foreground stellar contamination (red crosses), as simulated with the TRILEGAL models. The simulations include incompleteness effects and photometric errors in their color-magnitude distribution. We point out that the drawn distribution is only one random realization of the Galactic foreground model. On the left side of each diagram, representative $1\sigma$ photometric errorbars (as derived from artificial star tests) are shown. }}
 \label{cmds}
\end{figure*}

 Before discussing the features of our CMDs, we briefly discuss the Galactic foreground contamination issue. To estimate it, we use the TRILEGAL models \citep{girardi05}. Their web form\footnote{http://stev.oapd.inaf.it/cgi-bin/trilegal.} allows one to perform simulations of the Galactic population star counts in different directions and considering several free parameters (e.g., initial mass function, or IMF, extinction, thin and thick disk shape). For our simulations, we include exponential thin and thick disks plus oblate halo and bulge components. We point out that each of the performed simulations is a random realization of the theoretical model and not a unique representation of the Galactic foreground. At the sky position of the target objects, and within the ACS field of view ($\sim 3 \times 3$ arcmin), there are between 170 and 260 foreground dwarf stars (depending on the galaxy considered and using a Salpeter IMF). One has to carefully compare these numbers to what is seen in the CMD of each galaxy. In Fig. \ref{fore} we show the CMDs for our target galaxies, and in Fig. \ref{cmds} we additionally report their Hess density diagrams to which we overplot the simulated Galactic foreground stars from TRILEGAL as red crosses. For each magnitude bin, we have randomly extracted a subsample of stars from the original simulation in order to account for the incompleteness of the observations; moreover, we have randomly computed for each of the simulated stars a photometric error (depending on its magnitude) and have accordingly shifted its original position in the color-magnitude space by this amount. The position of the plotted foreground stars should thus not be taken as an absolute value, but as a possible realization of the expected foreground contamination. Overall, the contaminant fraction is less than $10\%$ for all of the galaxies and is only affecting the red part of the CMDs. However, when looking at the strips with magnitudes $23\lesssim I \lesssim25$ and colors in the range $1<V-I<1.3$ (red supergiant region, see description below), the contaminant fraction of foreground stars in this region is much higher, with values of $\sim22\%$, $\sim20\%$ and $\sim25\%$ for ESO443-09, ESO381-18 and ESO444-78, respectively. Also the asymptotic giant branch region appears to be contaminated, but due to the higher number of stars in this stage with respect to the red supergiant phase, that does not affect seriously our subsequent analysis. For IC4247 and ESO381-20, the CMD is overall well populated, and the fraction of Galactic foreground objects in the red supergiant region rather small ($5\%$ and $2\%$ respectively).

We thus decide to perform a statistical foreground subtraction over the whole CMD for the three galaxies where the contaminant fractions are not negligible. The decontamination process for the other two galaxies would leave all of our results unchanged because the contaminants fractions are very low, so we decide not to perform it in these cases. We adopt the following method: we first randomly extract a subsample of objects from the list of foreground stars simulated by the TRILEGAL models, in order to account for the incompleteness of our observations (which differs for different color and magnitude ranges). Then, we consider the observed galaxy stars that are found in a circle around each simulated foreground object, with a radius equal to two times the error in color at that position in the CMD. If more than one observed star is found within that circle, we simply subtract one of them randomly. We then use these decontaminated CMDs as an input for the SFH recovery process.

We also try an alternative approach to account for the foreground decontamination of the most affected galaxy in our sample, namely ESO444-78. Instead of subtracting the foreground stars in advance, we model the contaminant component during the SFH recovery process (see Sect. \ref{sfh_sec}). To do this, we first simulate a total of 10 random realizations of the foreground model for the considered galaxy, and treat them as a probability distribution in the color-magnitude space. During the SFH computation, we then let the code look for the best fit to the observed CMD for a model that contains dwarf galaxy stars plus Galactic foreground stars. In this way, we are statistically treating the foreground, as opposed to choosing one random realization of it. This test shows that the SFHs recovered with both methods are comparable to each other within the errors. The main resulting parameters (e.g., average SFR and metallicity) are consistent within the errorbars, and the only noticeable difference is observed for the SFR in the past $\sim40$ to 100 Myr. This is approximately the age range covered by red supergiant stars, which as stated above are found in the CMD region that has the highest contaminants fraction. In the case of ESO444-78, which has a very low SFR from 40 to 100 Myr ago as measured by main sequence stars, many of the bright red stars are better matched by scaling the foreground than by increased star formation in the galaxy itself. The net result is that when the foreground is explicitly modeled with the SFH recovery code, the derived SFR is a factor of $\sim2$ lower in this age range than what was found when the foreground was statistically subtracted based on a single TRILEGAL model simulation. This is a worst-case scenario because ESO444-78 has the smallest number of bright main sequence stars in our sample, giving the red supergiant/foreground region disproportionate influence over the derived SFH. For ages $\gtrsim100$ Myr, systematic effects due to the treatment of foreground modeling are negligible. Since for ESO444-78 we find a change of a factor of 2 in the SFR from  $\sim40$ to 100 Myr ago, we conclude that in this age range the errorbars resulting from the SFH recovery could be underestimated by at most this amount.

%________________________________________________________________

\section{Color magnitude diagrams} \label{cmd_sec}

The CMDs for our five target galaxies (ordered by absolute magnitude) are presented in Fig. \ref{fore}, and we additionally show their Hess density diagrams in Fig. \ref{cmds}. Also shown on the left-hand side of each diagram are the mean photometric uncertainties for each magnitude bin, as derived from the artificial star tests. For instance, the $1 \sigma$ error is $\sim0.1$ mag in magnitude and $\sim0.15$ mag in color at different $I$-band magnitudes for different galaxies (namely at $I=25.50$ for ESO443-09, $25.30$ for ESO381-18, $25.45$ for ESO444-78, $25.30$ for IC4247, and $25.35$ for ESO381-20). These error limits will be used in Sect. \ref{maps_sec} to define different stellar subsamples.

The CMDs of all of the galaxies exhibit the following evolutionary stages:

\begin{itemize}

 \item{upper main sequence (MS): a blue plume of main sequence candidates (and blue helium-burning stars) is found in the color range of $-0.5\leq V-I \leq0$ and at magnitudes of $I\geq22$ (Fig. \ref{cmds}). The youngest and most massive stars detected in our CMDs have estimated ages of $\sim10$ Myr;}

\item{helium-burning stars: massive stars with ages from $\sim20$ to 500 Myr are visible in the CMDs in the so called ``blue-loop'' phase, burning helium in their core. These evolved supergiants comprise blue ($V-I\sim0$ to 0.5) and red ($V-I\sim1$ to 1.3) supergiants. The latter are however not prominent for the galaxies in which the star formation was low some $\sim100$ to 500 Myr ago (ESO443-09, ESO381-18 and ESO444-78);}

 \item{asymptotic giant branch (AGB): lower mass stars with ages between $\sim0.1$ and 8 Gyr are found along the red giant branch and above its tip. The latter (luminous AGB stars) are well visible at colors of $1.3\leq V-I\leq2.5$, except for ESO443-09, which contains very few of these objects;}

 \item{upper red giant branch (RGB): these are evolved stars with colors in the range $V-I\sim1$ to 1.8, and indicative of intermediate-age and old ($\geq 1-2$ Gyr) low-mass stars which have not yet commenced core helium-burning. Up to $\sim20\%$ of the stars along the apparent upper red giant branch are in fact faint AGB stars in the same age range as the red giant branch stars \citep{durrell01}.}

\end{itemize}

\begin{figure}
 \centering
  \includegraphics[width=9cm]{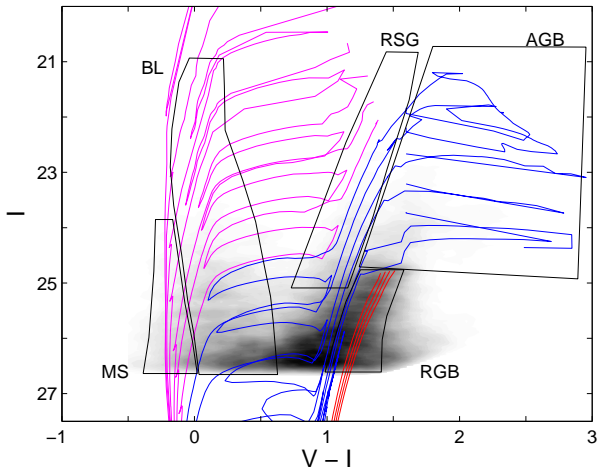}
 \caption{\footnotesize{Hess density diagram of the dwarf irregular galaxy ESO381-20. Overlaid are Padova stellar isochrones, with a fixed metallicity of Z=0.0008 and varying ages. Ages of 4, 8, 15, 20, 35, 55 and 85 Myr (proceeding from the blue to the red part of the CMD) are drawn in magenta, ages of 130, 200, 350, 550, 800 Myr and 1.3 Gyr are in blue, and ages of 4, 7, 10 and 14 Gyr are in red (see text for details). Also plotted are selection boxes that separate different evolutionary stages, as indicated.}}
 \label{isos}
\end{figure}

To help the eye recognize the various features, we plot again the Hess diagram for the galaxy ESO381-20 in Fig. \ref{isos} and overlay Padova isochrones. We choose a metallicity of Z=0.0008 (which roughly corresponds to [Fe/H]$\sim-1.4$, assuming Z$_\odot$=0.019), since this is the best-fit average metallicity resulting from our synthetic CMD modeling (see Sect. \ref{sfh_sec}). The isochrone ages range from 4 Myr to 14 Gyr (from the blue to the red side of the CMD). We draw the ``young'' ages (4 to 85 Myr, encompassing upper MS and most of the blue helium-burning stars) in magenta, the ``intermediate'' ages (130 Myr to 1.3 Gyr, clearly showing where most of the red helium-burning stars and the luminous AGB stars lie) in blue, and the ``old'' ages (4 to 14 Gyr, for which only the upper red giant branches and luminous AGB stars are visible at this photometric depth) in red. These isochrones were truncated at the tip of the red giant branch omitting the thermally pulsing AGB stars, to avoid overcrowding in the plot. We point out that the previous color-coding of ``young'', ``intermediate'' and ``old'' ages is arbitrary, and simply meant to illustrate the diverse features of the CMD, which were described above.

In Fig. \ref{isos} we also show the selection boxes that separate different evolutionary stages for ESO381-20, which are used in Sect. \ref{maps_sec}. We select different evolutionary stages in the CMDs following the stellar sequences for each individual galaxy and referring to the Padova isochrone models. The stellar content of each galaxy is divided in: RGB, MS, blue helium-burning stars (or blue loop, BL), red helium-burning stars (or red supergiants, RSG), and luminous AGB. We stress that the upper RSG region is not well reproduced by the models, in the sense that the observed stars are redder than predicted \citep[e.g.][and references therein]{ubeda07}, and we thus follow the stellar sequence on the CMD to define our selection box. We will come back to this point in Sect. \ref{sfh_sec}.

The presence of many different evolutionary stages clearly indicates the presence of a prolonged star formation, and each one represents a range of stellar ages. Depending on the number of stars that are found in each stage and depending on the time resolution that the CMD offers, it is possible to quantitatively constrain the SFR of a galaxy during its past history. In the following analysis, our purpose is to derive the SFH of the target galaxies, making use of a code written by one of the authors (A. A. Cole).

%________________________________________________________________

\section{Star formation histories} \label{sfh_sec}

\begin{table*}
 \centering
 \caption{\footnotesize{Star formation parameters derived for our sample of galaxies.}} 
 \label{ressfh}
\begin{tabular}{lccccccccc}
\hline
\hline
Galaxy&$<SFR>$&$b_{100}$&$b_{500}$&$b_{1G}$&$b_{14G}$&$f_{1G}$&$f_{4G}$&$f_{14G}$&$<$[Fe/H]$>$\\
&($10^{-2}$M$_{\odot}$yr$^{-1})$&&&&&&&&(dex)\\
\hline
\object{ESO443-09, KK170}&$0.08\pm0.07$&$6.48$&$3.02$&$0.91$&$1.01$&$0.11$&$0.06$&$0.83$&$-1.46\pm0.14$\\
\object{ESO381-18}&$0.35\pm0.26$&$1.71$&$0.93$&$1.91$&$0.95$&$0.13$&$0.11$&$0.76$&$-1.40\pm0.13$\\
\object{ESO444-78, UGCA365}&$0.63\pm0.36$&$0.66$&$0.48$&$0.40$&$1.07$&$0.03$&$0.11$&$0.86$&$-1.37\pm0.21$\\
\object{IC4247, ESO444-34}&$0.61\pm0.45$&$1.04$&$1.06$&$0.70$&$1.07$&$0.05$&$0.32$&$0.63$&$-1.37\pm0.21$\\
\object{ESO381-20}&$0.70\pm0.48$&$4.36$&$3.41$&$1.76$&$0.92$&$0.15$&$0.06$&$0.79$&$-1.45\pm0.17$\\
\hline
\end{tabular}
\tablefoot{
The columns are the following: column (1): name of the galaxy; (2): average SFR over the whole lifetime; (3): $b_{100}$; (4): $b_{500}$; (5): $b_{1G}$; (6): $b_{14G}$; (7): $f_{1G}$; (8): $f_{4G}$; (9): $f_{14G}$; and (10): average metallicity over the whole lifetime. The $b$ parameter is the ratio of star formation rate over the indicated time period (0.1, 0.5, 1 and from 1 to 14 Gyr respectively) to the average star formation over the whole lifetime; the $f$ parameter is the fraction of stars born in a certain time interval ($0-1$, $1-4$ and $4-14$ Gyr). In a hypothetical galaxy with constant star formation rate, $f_{1G}=0.07$, $f_{4G}=0.36$ and $f_{14G}=0.57$.
}
\end{table*}

We used synthetic CMDs based on published theoretical isochrones to constrain the SFRs as a function of time over the recent history of the galaxies. We used the code developed by Cole \citep{skillman03, cole07} to find the most probable SFH based on minimizing the difference between sets of synthetic CMDs and the data.  This approach is widely used in the interpretation of the CMDs of resolved stellar populations \citep[e.g.,][and references therein]{gallart05, tolstoy09}. While the method is most powerful when the CMD includes the MS turnoff of the oldest stellar populations, robust results at lower precision can be derived even from relatively shallow data. For example, the inferences of \citet{tolstoy98} about the Local Group dwarf Leo~A (DDO~69) were confirmed with increased precision by the much deeper data of \citet{cole07}. The method adopted here is similar in its general approach to the methods used in other recent studies of dwarf galaxies in the Local Volume, e.g., the M81 group \citep{weisz08}.

The SFH-fitting code starts from a set of isochrones interpolated to a fine grid of age and metallicity in order to create synthetic CMDs with no artificial gaps between points. The synthetic CMDs are then binned in age and metallicity to increase computational tractability and conform to the information content of the data; this typically results in age/metallicity bins that are nearly evenly spaced in log(age) and log($Z$). The CMDs are divided into a regular color-magnitude grid, and the expectation value of the number of stars expected in each grid cell for a SFR of 1 M$_{\odot}$ yr$^{-1}$ is computed from the isochrones. There are several input parameters that must be fixed or fit in order to find physically meaningful solutions with good efficiency. Among the most important of these are the distance modulus $(m-M)_{0}$, reddening $E(B-V)$, and reddening law (expressed as a single-parameter R$_{\lambda}$ relation, \citealt{cardelli89}). Once an isochrone set is chosen, the density of stars along the isochrones is altered by the IMF, the frequency of binary stars, and the mass function of binary companions.

The isochrone is then convolved with the color and magnitude errors and incompleteness of the dataset in order to allow a direct comparison between the models and data. During the fitting process, linear combinations of the individual synthetic CMDs are added to find the composite CMD that best matches the observed data. Because many cells of the CMD grid contain relatively few stars, a maximum likelihood test based on the Poisson distribution \citep[][]{cash79, dolphin02} is used instead of a $\chi^2$ statistic. Because there may be 50--100 individual synthetic CMDs considered, a direct search of parameter space is not feasible; however, the problem is well-suited to a simulated annealing approach in which an initial guess at the SFH is randomly perturbed and the increase or decrease in log-likelihood is calculated for each perturbation. Perturbations that increase the likelihood are retained while those that worsen the fit are usually rejected. The simulated annealing formalism prevents the code from being stuck in local minima, of which there are many. Prior to the perturbation of each trial solution, the SFRs are transformed using an arcsin function in order to ensure that only non-negative SFRs are sampled.

We adopted specific choices motivated by the best available information, choosing not to attempt to solve for parameters that were otherwise constrained (e.g., distance modulus), or about which our dataset contains little information (e.g., binary fraction, IMF). We used the Padova isochrones \citep{marigo08} with circumstellar dust around cool stars from \citet{groenewegen06}, combined with the IMF of \citet{chabrier03}. This IMF is identical to the Salpeter power law above 1 M$_\odot$ but is log-normal for low-mass stars. Based on the statistics in \citet{duque91} and \citet{mazeh92}, one third of the stars are single and the rest have a companion star drawn from the same IMF as the primary. Distances based on the TRGB method are taken from \citet{kara07}, and reddenings from the extinction maps of \citet{schlegel98}. We did not {\it a priori} assume any internal reddening within the target galaxies. However, our fit procedure accomodates the possibility by allowing the reddening value to vary slightly. To understand the possible effects that internal reddening might have on our results, we perform some tests on the galaxy IC4247 (see Sect. \ref{sec_4247}). The fitting code does not take differential reddening into account, but \citet{cignoni10}, in their review of SFH modelling techniques, show that the effects of differential reddening on SFH reconstruction are typically minor and of less importance than proper constraints on the foreground and average reddening. Because we are dealing with metal-poor dwarf galaxies, the amount of dust is expected to be generally low.

\begin{figure*}
 \centering
  \includegraphics[scale=0.57]{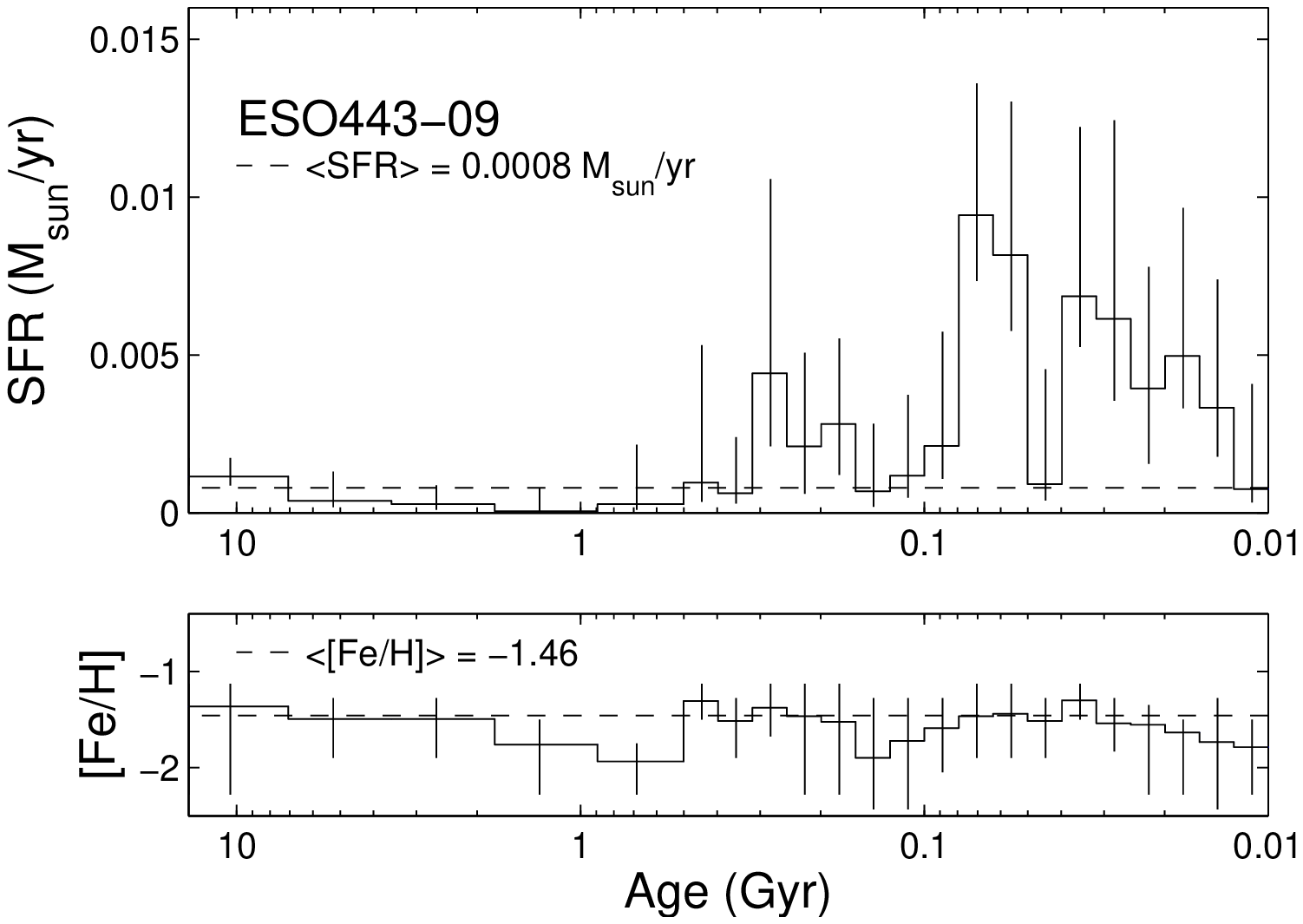}
  \includegraphics[scale=0.57]{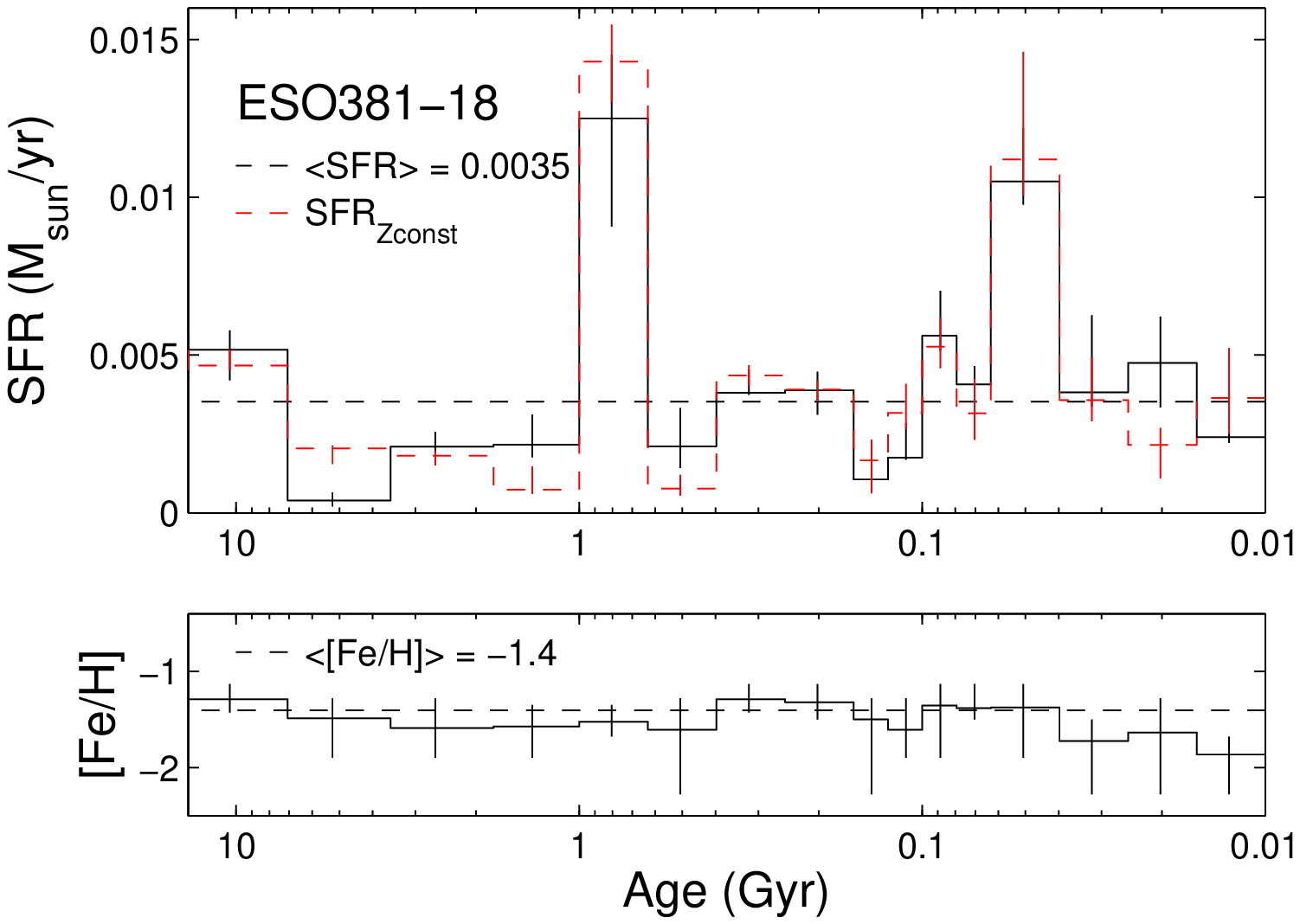}
  \includegraphics[scale=0.57]{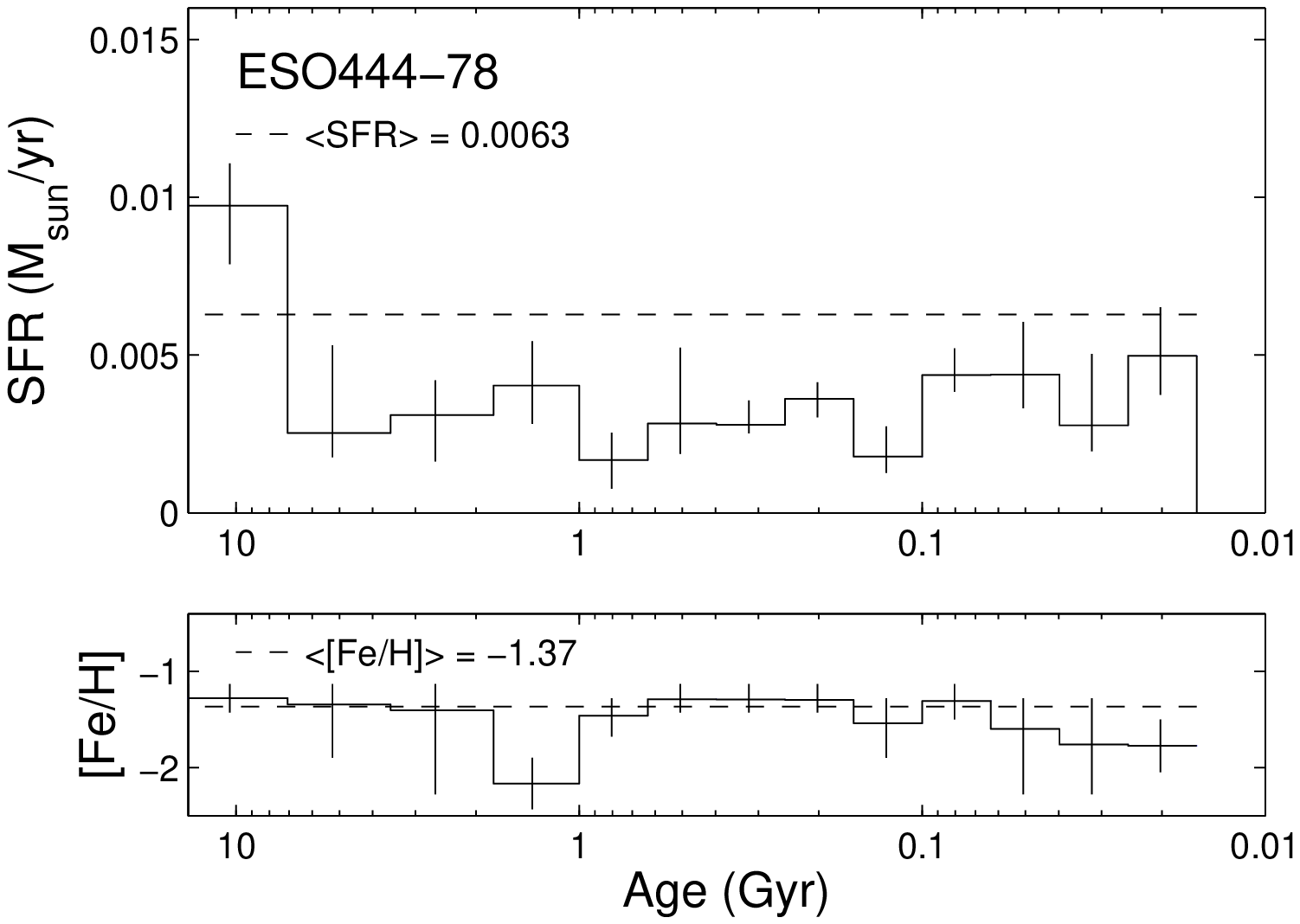}
  \includegraphics[scale=0.57]{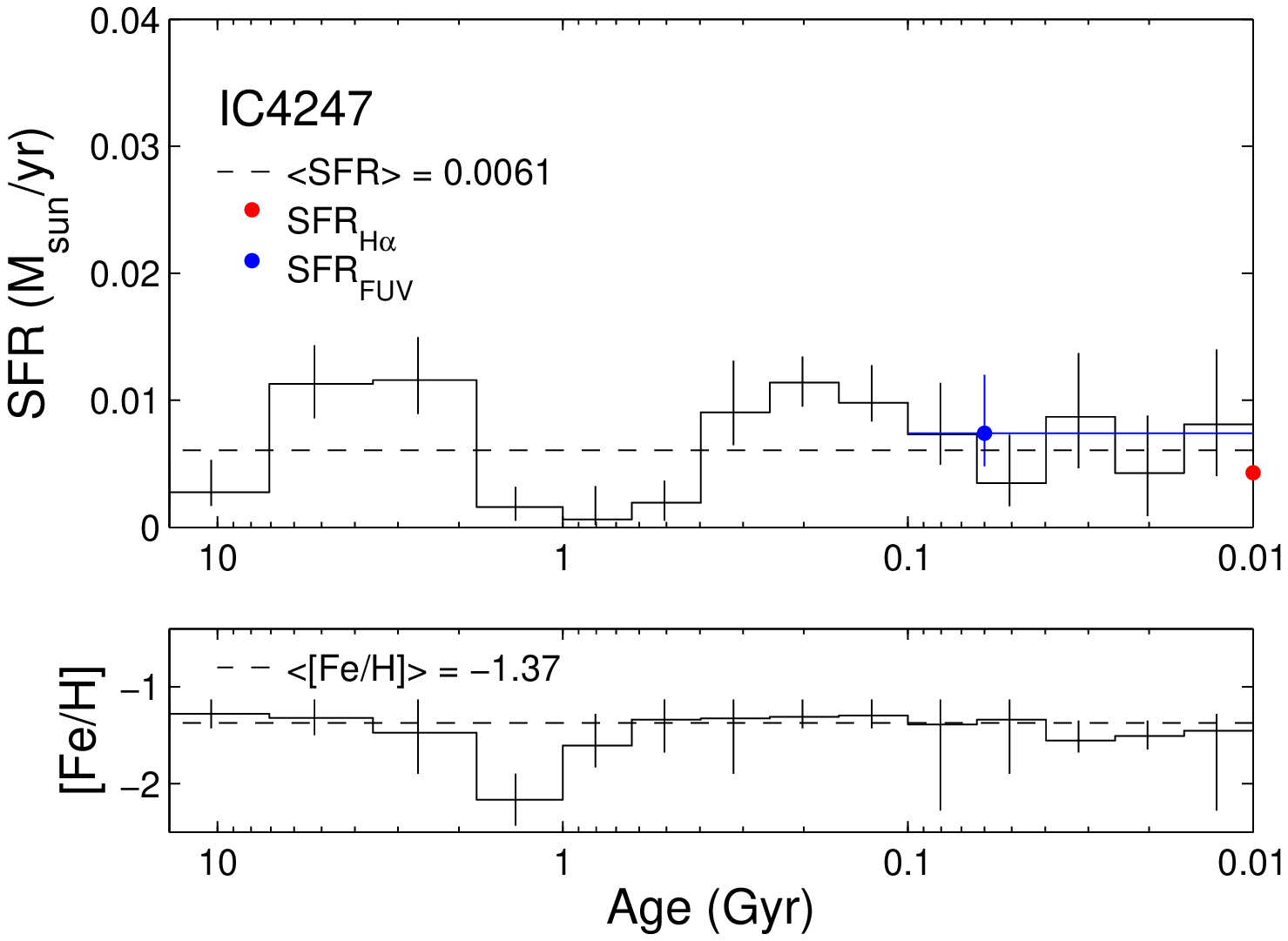}
  \includegraphics[scale=0.57]{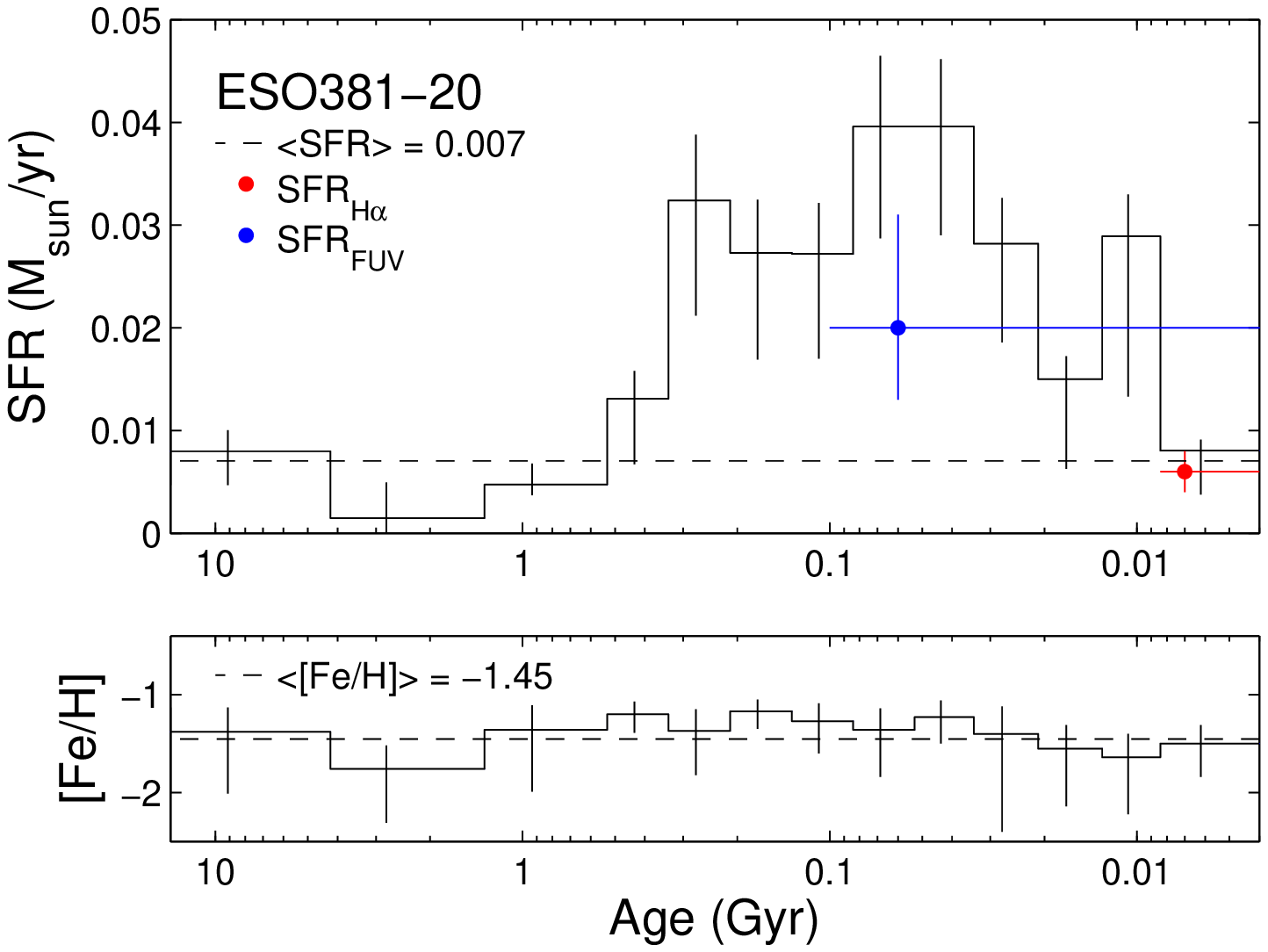}
 \caption{\footnotesize{\emph{Upper panels}. Star formation histories derived for the five studied galaxies (ESO443-09, ESO381-018, ESO444-78, IC4247 and ESO381-20, ordered by increasing absolute magnitude). For each galaxy, the star formation rate as a function of time is plotted, with the oldest age being on the left side and the most recent age bin on the right edge of the (logarithmic) horizontal axis. Note that the size of the age bins is variable due to the different amount of information obtainable from each CMD for each stellar evolutionary stage. ESO381-20 is the only galaxy in this sample for which the SFR could be derived for ages younger than 10 Myr. Also note the different vertical axis scales. The black dashed line indicates the mean star formation rate over the whole galaxy's lifetime. The red dashed line for ESO381-18 indicates the alternative star formation history solution obtained when restricting the metallicity range in the parameter space (see text for details). The red dots for IC4247 and ESO381-20 report the current star formation rate derived from H$\alpha$ observations described in the literature, while the blue dots indicate the recent star formation rate derived from the FUV (see text for references). \emph{Lower panels}. Metallicity as a function of time, with the same axes convention as above. The black dashed line represents the mean metallicity over the galaxy's lifetime. Note that the metallicity evolution is poorly constrained.}}
 \label{sfhsnew}
\end{figure*}

No age-metallicity relation is assumed, but where possible we use literature values to constrain the present-day metallicity estimated from nebular oxygen abundances (available for IC4247 and ESO381-20, see below for details). In these cases, the highest metallicity considered was taken to be the upper bound on metallicity from the nebular abundances. The full range of metallicities considered was $Z = 0.0001, 0.00024, 0.0004, 0.0006$, and $0.001$. Higher metallicities can in general be ruled out because of the mismatch in color of the evolved stars (RGB). Moreover, allowing a larger range in metallicities would increase the computational time and would result in higher uncertainties in the metallicity values. As a test, we choose isochrones with metallicities as high as $Z = 0.005$ for one of the target galaxies (ESO381-20), to check whether this would change the results of our SFH recovery. The average SFR obtained in this case is just slightly lower (the difference is $\sim 10\%$ in M$_\odot$ yr$^{-1}$) than the original value, while the metallicity is higher by $\sim0.2$ dex, both being still consistent with the original values within the errorbars. The only notable difference lies in the shape of the recent SFH ($\lesssim 500$ Myr): there is a shift of a few $\sim 10^7$ yr in the age of the youngest episodes of star formation, where the age shifts around due to a combination of age-metallicity degeneracy and contravariance between adjacent age bins. This is however still comparable to the original SFH within the errorbars, and also the star formation parameters that we compute (see below) remain the same. Metallicity ranges at a given age are not ruled out by our data, and isochrones of multiple metallicities typically contribute to the total star formation in each age bin.

Finally, no information is available regarding the detailed abundance patterns, or even approximations such as average [$\alpha$/Fe] values, so no attempt is made to model this parameter. Little is known about [$\alpha$/Fe] values in dwarf irregular galaxies and in particular we have no measurements for those beyond the Local Group. We can however assume that galaxies outside the Local Group follow similar trends to the Local Group ones. The existing studies for the Local Group tend to suggest that for younger populations and/or higher metallicities the [$\alpha$/Fe] ratios are around solar values, while for very old populations and/or low metallicities the [$\alpha$/Fe] ratios tend to be similarly high as in the Galactic halo \citep[see, e.g.,][]{venn01, venn03, taut07, tolstoy09}. If the [$\alpha$/Fe] ratios at those early ages were higher than they are now, then this would affect the RGB stars in our CMDs. For example, at the TRGB and at a fixed metallicity of [Fe/H] $=-1.5$, assuming a value of [$\alpha$/Fe] $\sim +0.3$ dex would mimic a higher age ($\sim2-3$ Gyr for ages $\gtrsim7$ Gyr, see Dartmouth isochrones, \citealt{dotter08}). On the other hand, since we can only poorly resolve ages older than $\sim7$ Gyr, which are the ones where a departure of [$\alpha$/Fe] from solar values might be present, considering different [$\alpha$/Fe] values would not alter our current result. Since the uncertainties in our derived SFHs are the largest for ages older than a few Gyr, the only effect of adding [$\alpha$/Fe] as a free parameter in the maximum likelihood process would be to make the errobars larger for those time bins. Moreover, the Padova isochrones that we adopt throughout our study do not model variations of [$\alpha$/Fe] values.

When the code indicates that the global maximum likelihood fit has been reached, errorbars on the SFRs are computed by perturbing the SFR of each age bin in turn, holding the rate fixed at the perturbed value and finding the new best-fit solution. When the new best fit is no longer within 1$\sigma$ of the likelihood of the global best fit, the change in the SFR of the bin under consideration is taken to be the errorbar on the SFR. The value is then reset to the best-fit value and allowed to vary again, and the process is repeated for the next bin until all values have both an upper and a lower errorbar computed.  Typically the SFRs of adjacent bins are highly anticorrelated, because the total number of stars is fixed, so that a reduction in SFR at one age forces the synthetic galaxy to make up the difference at older and/or younger ages. These changes then propagate into the metallicity distribution because an increase in the metallicity of an isochrone can partially mask a decrease in age.

The code output is a table of SFRs and mean metallicities as a function of age in the user-defined age bins. From these, the synthetic CMDs can be reconstructed and compared to the data for a visual impression of the fit quality. The absolute value of the maximum likelihood depends sensitively on the gridding scheme for the CMD, the treatment of contaminating data points not in the isochrones (e.g., Galactic foreground stars), and the matching of a few stars in bright but sparsely populated (and possibly poorly modelled) phases of evolution. However, the most likely SFH to match the CMD is driven by the distribution of the most populous cells in the gridded CMD and is robust as long as the incompleteness is carefully modelled. The best-fit SFHs and age-metallicity relations are shown in Fig. \ref{sfhsnew}.

The first thing to point out is that we plot the SFHs with a logarithmic age axis. This is done because the age resolution decreases with increasing age. The age bins will be broader when there is less information available from the CMD, and finer where there is more. For example, in Fig. \ref{isos} we see that the oldest main sequence turnoff detectable in our data is visible for stars no older than $\sim50$ Myr ($I\sim26$). For the recent SFH we can then rely on the upper MS and the bright helium-burning phases. The CMDs do not reach the horizontal branch (which at these distances would be expected at an $I$-band magnitude of $\sim28$). Also, the red clump is not recognizable in our data. This means that for ages older than $\sim1$ Gyr it is very difficult to put any firm age constraint on star formation episodes from the evolved population alone. We can put constraints on the intermediate-age SFR ($\sim1$ to $\sim9$ Gyr) by looking at luminous AGB stars, but we are not able to resolve any bursty episode of star formation at these ages, and so the time bins become very large as age increases. Each bin size is thus considered as a horizontal errorbar. 

As mentioned in the previous Section, theoretical models are usually not reproducing perfectly the supergiant helium-burning phases due to the difficulties in robustly modeling these evolutionary stages \citep[e.g.,][]{maeder01, dohm02b, levesque05}. The models tend to predict colors bluer than observed for RSGs, which would result in metallicities biased towards higher values, but at the same time the observed color of the blue helium-burning stars prevents this effect. In our case this discrepancy does not heavily influence the resulting SFHs. The number of stars in these phases is small for all target galaxies (so their weight in the SFH recovery process is lower) except for IC4247 and ESO381-20, but the latter already have metallicity constraints from HII regions.

\begin{figure}
 \centering
  \includegraphics[width=7cm,height=6cm]{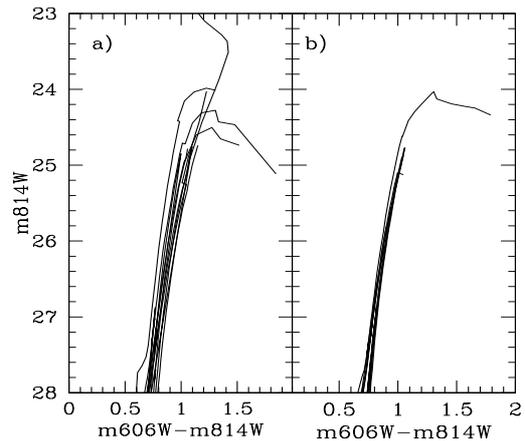}
 \caption{\footnotesize{Padova stellar isochrones with a range of ages (1, 2, 4, 8 and 14 Gyr from left to right in each panel), shown for a combination of HST/ACS filters. \emph{Panel a).} The metallicity is set to be constant ($Z = 0.001$). \emph{Panel b).} The metallicity has an initial value of $Z = 0.0004$ at 14 Gyr, and increases as age decreases ($Z = 0.0006$, $0.001$, $0.0015$, and $0.0024$, respectively). Note the difference in width of the RGB.}}
 \label{agemet}
\end{figure}

The vertical errorbars for the star formation shown in the upper panels of Fig. \ref{sfhsnew} come from the maximum likelihood process, and extend to all of the values that produce solutions within $1\sigma$ from the best-fit star formation value as explained above. In the lower panels we also report the evolution of metallicity as a function of time. Unfortunately, due to the age-metallicity degeneracy for the old stars of the RGB and due to the limited amount of information for younger ages, it is very difficult to constrain the metallicity for our target galaxies. We thus underline that the mean metallicity values derived are reliable, while the evolution with time and potential metallicity spreads cannot be firmly constrained from the available data. However, we can at least rule out the presence of a strong metallicity increase with time by looking at the width of the RGB. In Fig. \ref{agemet} we give an example (using Padova isochrones) of the combined effect of age and metallicity on the width of the RGB. A constant value of metallicity coupled with decreasing age would produce a much broader RGB than a case where the metallicity is strongly increasing with decreasing age \citep[age-metallicity degeneracy, see also][]{worthey99}. From our observed CMDs we can thus exclude such a strong trend for metallicity, but we cannot distinguish between a scenario with constant metallicity and one with a mild evolution with time. We note that such mild metallicity evolution with a fairly flat age-metallicity relation has also been found for a few Local Group dwarfs using photometry and stellar spectroscopic metallicities \citep[e.g.][]{koch07a, glatt08b}. Finally, the errorbars in metallicity show the range, in each age bin, from the $10^{th}$ to the $90^{th}$ percentile from the best-fit SFH.

We note however that there is disagreement between our derived [Fe/H] values and previous literature works (\citealt{kara07}, \citealt{sharina08}), in the sense that our values are higher for all of the target dwarfs. In those studies, the mean metallicity of the target galaxies was computed through the empirical formula by \cite{lee93} considering the mean $V-I$ color of stars on the upper red giant branch. The aforementioned formula is valid only for predominantly old populations, while in our target galaxies there is clear evidence for intermediate-age stars. These younger stellar components will bias the old RGB in the sense that it will be highly contaminated by intermediate-age RGB stars, which have bluer colors than old RGB stars at a fixed metallicity and luminosity \citep[see, e.g.,][]{cole05}. When adopting the \cite{lee93} formula, RGB stars younger than $\sim10$ Gyr will thus be considered to be old, metal-poor RGB stars, and the resulting metallicity may be underestimated by up to 0.5-1 dex. In addition, faint AGB stars along the RGB locus can bias the derived metallicities to erroneously lower values.

As an additional test, we compute the total number of stars formed during the whole galaxy's lifetime from our SFH, and compare it to estimates of the stellar mass coming from the total luminosity and an assumption for the stellar mass-to-light ratio. The latter has normally a value of $\sim1$ for gas-rich dwarf galaxies \citep[e.g., ][ and references therein]{banks99, read05}. \citet{cote00} analyze dwarf irregulars in the CenA and Sculptor groups. We have one galaxy, ESO381-20, in common with their sample. Looking for the best-fit stellar mass-to-light ratio from their HI rotation curves, they derive values larger than 1 ($\sim2$ to $\sim4$) for some of the galaxies including ESO381-20, so we decided to additionally assume a stellar mass-to-light ratio of 2 when computing stellar masses for our sample. We find good agreement between the two different estimates, and we list the exact values in the individual galaxies' description. 

To better understand the SFH, we also present some parameters to facilitate the comparison of the results among individual galaxies and to other literature studies \citep[e.g., ][]{scalo86, weisz08}. We compute the ratio of the SFR in a certain time period to the average SFR during the galaxy's lifetime ($b_{100}$ over the last 100 Myr, $b_{500}$ over the last 500 Myr, $b_{1G}$ over the last 1 Gyr, and $b_{14G}$ for ages older then 1 Gyr ago). We also derive the fractions of stars formed from $0-1$ Gry ($f_{1G}$), from $1-4$ Gyr ($f_{4G}$) and from $4-14$ Gyr ($f_{14G}$), to assess the efficiency of the star formation process in different time periods. These parameters would have the following values if a constant SFR were adopted: $f_{1G}=0.07$, $f_{4G}=0.36$ and $f_{14G}=0.57$. The values that we derive for these parameters, together with the lifetime average metallicity values, are reported in Tab. \ref{ressfh}. 

We now comment separately the results obtained for each individual galaxy.

\subsection{ESO443-09, KK170} \label{}

The faintest and least dense galaxy in our sample is ESO443-09. This object also contains the smallest amount of HI gas ($M_{HI}\sim10^{7}$M$_\odot$), and it is the most isolated one, having a tidal index of $-0.9$. Its distance from us is $5.97\pm0.46$ Mpc, and its deprojected distance from M83 is $\sim1.2\pm0.4$ Mpc (computed using the radial distance and the coordinates from \citealt{kara07}). ESO443-09 has not yet been studied in H$\alpha$. We detect only $\sim1900$ stars in its CMD (Fig. \ref{cmds}). ESO443-09 indeed contains very few massive MS stars and no evidence of currently ongoing star formation as judged from its CMD. There is, however, a considerable presence of BL stars, on top of the old component seen in the RGB; the RSG and an intermediate-age population (luminous AGB) are almost absent. Hence, overall this galaxy exhibits the properties of a typical transition-type dwarf \citep[e.g.][]{grebel03}.

The SFH derived for ESO443-09 and shown in Fig. \ref{sfhsnew} confirms that this galaxy has been almost constantly active in its early history, although with a low average SFR ($\sim0.0008\pm0.0007$M$_\odot$ yr$^{-1}$). From now on we adopt a standard $\Lambda$CDM cosmology with $t_0=13.7$ Gyr, $H_0=71$ km s$^{-1}$ Mpc$^{-1}$, $\Omega_{\Lambda}=0.73$ and $\Omega_{m}=0.27$ \citep{jarosic11}. If we assume that the galaxy was born $\sim13.5$ Gyr ago, the fraction of stars that had formed by 8 Gyr ago (which corresponds to $z\sim1.1$) and by 5 Gyr ago ($z\sim0.5$) is $\sim60\%\pm10\%$ and $\sim75\%\pm10\%$, respectively. This is just a rough estimate, since we cannot derive the age of the oldest stellar populations with these data. A fairly large fraction ($>10\%$) of stars was however formed in the last Gyr. In particular, the values of $b_{100}$ and $b_{500}$ (Tab. \ref{ressfh}) are quite high, showing a non-continuous episode of star formation from 10 to 200 Myr ago.

The metallicity is fairly constant within the errorbars. However, for this and for almost all of the other galaxies (except for ESO381-20), the present-day values appear to be slightly lower than the average ones. We will return to this issue in the next Subsection.

We compute the total stellar mass firstly from our SFH, which yields $1.1\times10^{7}$M$_\odot$. We note that this is not a precise estimate, since we are not taking into account the stars that died off due to their evolution. This implies that we are overestimating the stellar masses by $\sim20\%$. Starting from the $B$-band luminosity and assuming a stellar mass-to-light ratio of 1 or 2, we obtain values of $0.8\times10^{7}$M$_\odot$ and $1.6\times10^{7}$M$_\odot$, respectively, thus consistent with our SFH derivation.

\subsection{ESO381-18} \label{}

ESO381-18 is another rather isolated galaxy ($\Theta=-0.6$), found at a deprojected distance of $\sim1.2\pm0.1$ Mpc from M83 (computed using the radial distance from \citealt{kara07}). As for ESO443-09, neutral gas has been detected ($M_{HI}\sim3\times10^{7}$M$_\odot$), but there is no evidence of ongoing star formation (poorly populated upper MS, not observed in H$\alpha$). Also the BL and RSG phases are very sparsely populated, but a clump of AGB stars is visible in the CMD in Fig. \ref{cmds}, the latter containing a total of $\sim7300$ stars. ESO381-18 has a particularly high stellar density, with a peak stellar density of 909 stars per $0.1$ kpc$^{2}$ in its central regions.

This galaxy formed $55\%\pm3\%$ of its stars prior to 8 Gyr ago, while $72\%\pm3\%$ were at place by 5 Gyr ago. The SFR during the past $\sim100$ Myr and the past $\sim1$ Gyr (see $b_{100}$ and $b_{1G}$ in Tab. \ref{ressfh}) was higher than the average SFR value, which is $\sim0.0008\pm0.0007$M$_\odot$ yr$^{-1}$. The derived SFH in Fig. \ref{sfhsnew} suggests the occurence of two strong recent bursts, the first from 600 Myr to 1 Gyr ago and the second one, shorter, from 40 to 60 Myr ago.

\begin{figure}
 \centering
  \includegraphics[scale=0.23]{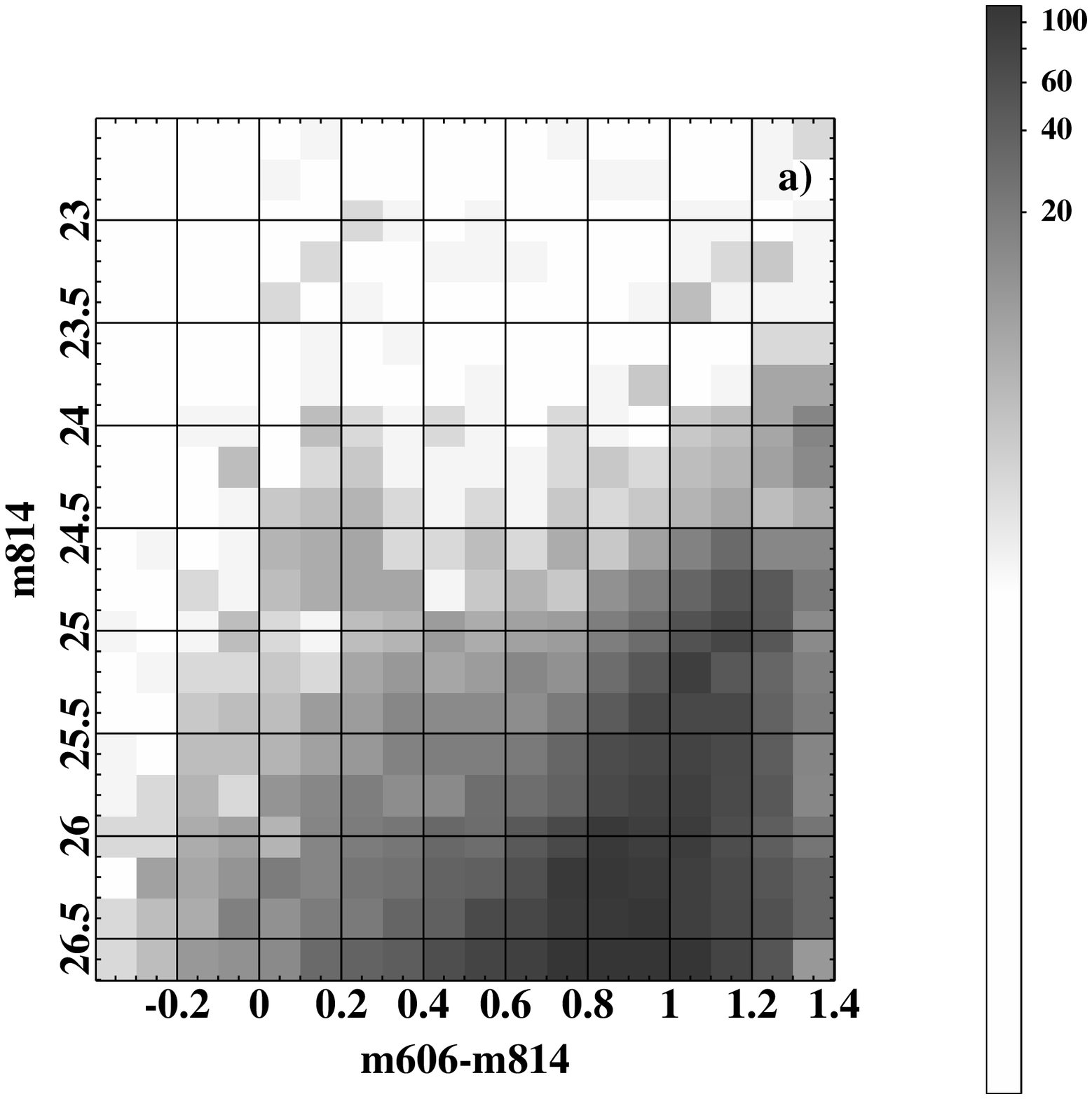}
  \includegraphics[scale=0.23]{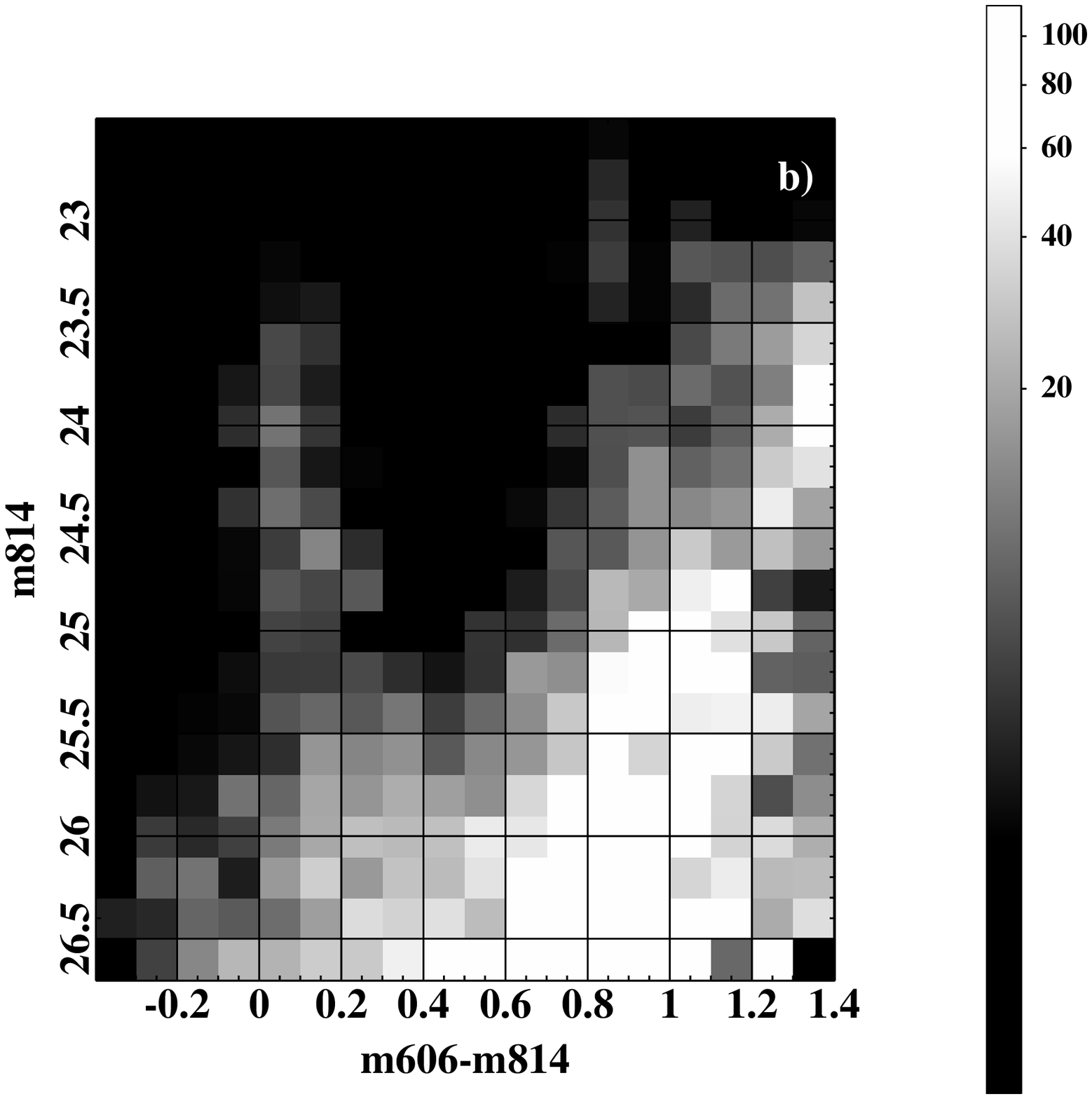}

\vspace{0.5cm}
  \includegraphics[scale=0.23]{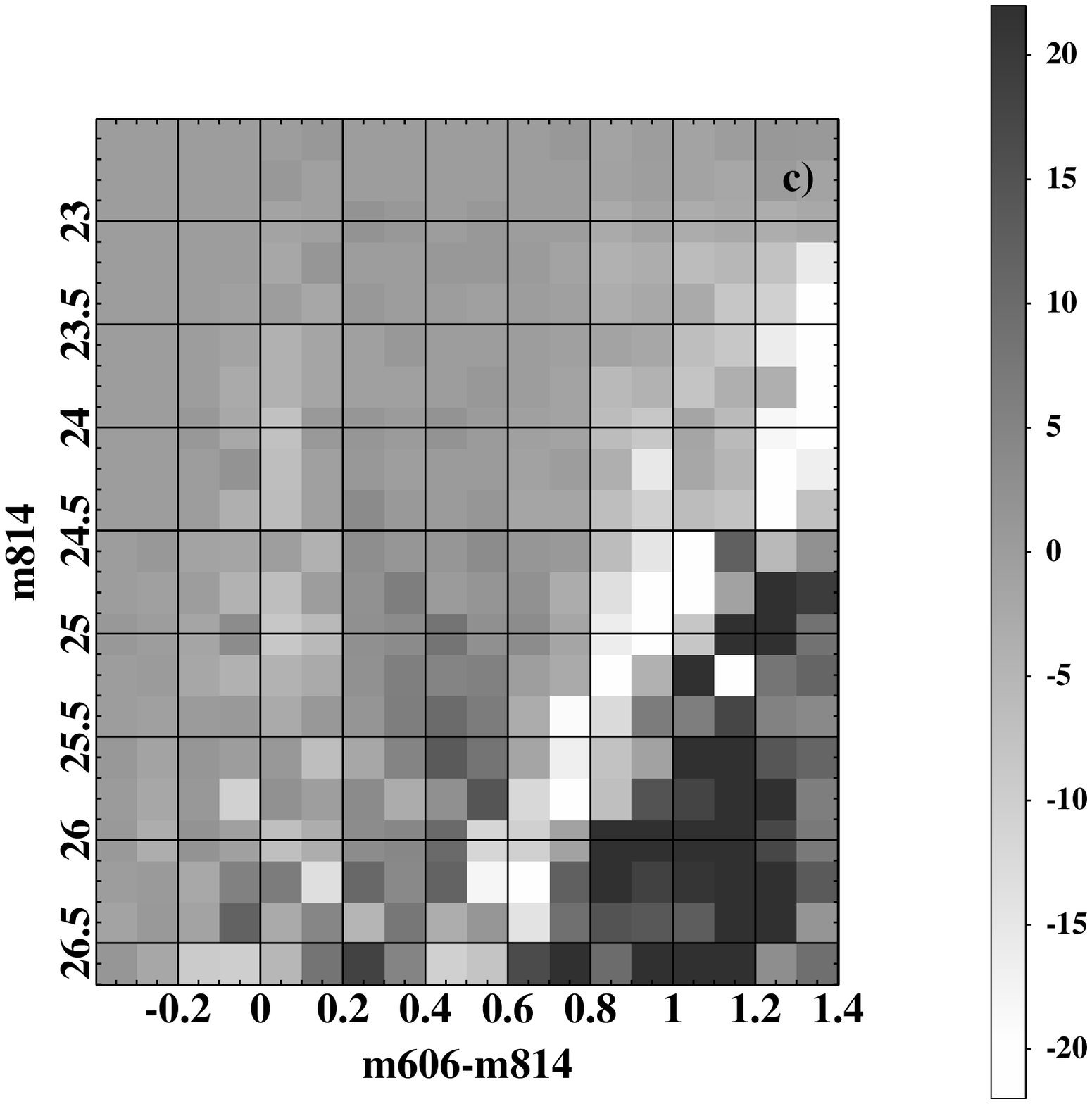}
  \includegraphics[scale=0.23]{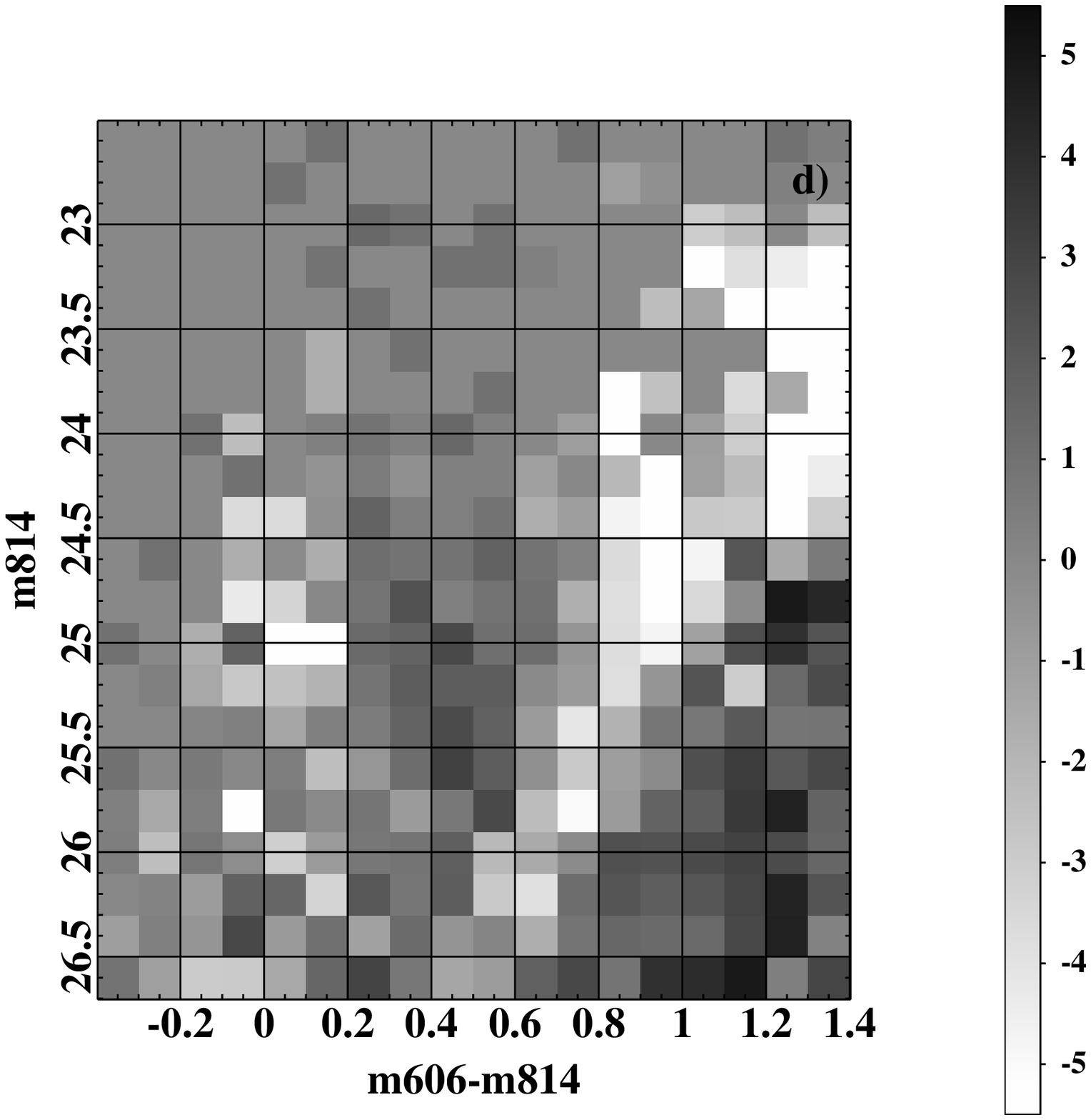}
 \caption{\footnotesize{Hess diagrams (displayed using ACS filters) for ESO318-18. The density values are listed along the colorbars. \emph{Panel a).} Data; the reported densities are in units of number of stars per bin (with a binsize of 0.10 mag in color times 0.20 mag in magnitude). \emph{Panel b).} Best-fit synthetic model, with a reverse color scale with respect to the data to facilitate comparison; units are number of stars per bin. \emph{Panel c).} Difference between data and best-fit model; units are number of stars per bin. \emph{Panel d).} Difference between data and best-fit model, weighted by the Poisson errors; units are proportional to the square root of the number of stars per bin.}}
 \label{38118pan}
\end{figure}

Some difficulties were met when deriving the SFH for this galaxy. In particular, the color of the old RGB and young BL populations and the TRGB were not well fitted by isochrones with the average metallicity we derive for this galaxy (see Tab. \ref{ressfh}), and with the reddening and distance modulus adopted. We thus first tried to slightly change the reddening and distance modulus. The best-fitting values found to reproduce the observed CMD were $E(B-V)=0.1$ instead of 0.06, and $(m-M)_{0}=28.56$ instead of $28.63\pm0.14$, which is still consistent with the original value within the errorbars. LEDA\footnote{http://leda.univ-lyon1.fr/.} reports an internal extinction value in the $B$-band of 0.42 mag due to the inclination of the galaxy, which may explain part of the discrepancy. However, the region above and redward of the RGB in the best-fit synthetic CMD does not match the observed one well. We plot the Hess diagram for the data and the best-fit model in Fig. \ref{38118pan} to show where the fit yields poor results. Note the bin size is big due to the low number statistics. The discrepancy is partly due to inconsistencies in the models, which cannot reproduce simultaneously the colors of RGB and RSG/BL stars at these low metallicities \citep[e.g.][]{ubeda07}, together with the poor number statistics and the possible presence of differential reddening.

Another point is that the metallicity seems to be decreasing with time (see Fig. \ref{sfhsnew}), and the problem is again related to the color of the RGB, which is not well fitted by the chosen set of isochrones. The metallicity value for the RGB (as high as $Z\sim0.004$ if we compare its color to models) is well above the values one would expect for such a small galaxy, also considering the low metallicity of the young populations we derive with the SFH recovery ([Fe/H]$\sim-1.4$, corresponding to $Z\sim0.0008$). Apart from the reddening issue, another plausible explanation could be an $\alpha$-element enhancement at early times, which would make the RGB redder. We perform the following test: we run the SFH recovery process again, but this time with a restricted range of metallicities ($Z=0.0004, 0.0006$ and 0.001). This will force the metallicity to vary very little, thus not giving us any information about the chemical evolution itself, but it will show how much the RGB color issue can affect our derived SFH. The results show a synthetic CMD that is almost identical to the first one, and the SFH (plotted in red in Fig. \ref{sfhsnew}) is comparable to our originally derived SFH within the errorbars. The second oldest time bin changes slightly to a higher SFR, but this age range is in any case quite uncertain, and the overall features (the two recent burst episodes) are still clearly distinguishable. Also the two most recent time bins show little change, but this is due to the higher metallicity imposed (with respect to the first model). We thus conclude once again that the derived SFH is fairly robust, but the constraints on the metallicity are weak.

Finally, we compare the stellar mass found from the SFH ($4.7\times10^{7}$M$_\odot$) with a mass estimate coming from the galaxy's luminosity and an adopted stellar mass-to-light ratio of 1 or 2. We find values of $2.2\times10^{7}$M$_\odot$ or $4.4\times10^{7}$M$_\odot$, respectively, thus again matching well the result of our SFH (even after correcting for the evolutionary effects).

\subsection{ESO444-78, UGCA365} \label{}

The dwarf irregular ESO444-78 is the closest one to M83 within our sample (only $\sim110\pm500$ kpc deprojected distance, again considering the values reported \citealt{kara07}), and is also currently located in the densest environment, with respect to the other targets of our study ($\Theta=2.1$). According to LEDA, also in this case there is a high internal extinction in the $B$-band due to the inclination (0.88 mag). As the previous two galaxies, ESO444-78 contains little neutral gas (see Tab. \ref{infogen}), but it was detected in H$\alpha$ \citep{cote09}. From the CMD alone ($\sim11400$ stars), the old stellar component seems predominant, while there are few, smoothly distributed stars in the young and intermediate-age phases, with a smaller concentration in the luminous AGB.

Indeed, ESO444-78 formed already $65\%\pm5\%$ of its stars more than 8 Gyr ago, and $82\%\pm5\%$ more than 5 Gyr ago. The fraction of the galaxy's total stellar content formed prior to 8 Gyr is the largest for this galaxy, compared to the others in the sample. The average SFR for ESO444-78 is $\sim0.0063\pm0.0036$M$_\odot$ yr$^{-1}$). For ages younger than 8 Gyr, ESO444-78 seems to have experienced an almost constant and low level of activity (see Tab. \ref{ressfh}), with only $3\%$ of its stars born in the last Gyr. The current SFR is estimated to be only $\sim0.000032$M$_\odot$ yr$^{-1}$ from its one HII region \citep{cote09}. We do not plot this value in the SFH of Fig. \ref{sfhsnew}, since we are not able to recover a significant SFR for the youngest time bin ($<10$ Myr). Also in this case, the metallicity seems rather constant, excluding the time bin between 1 and 2 Gyr ago, which has a lower value. This could be easily due to the fact that there is not much information for these ages in the CMD, so it may not be relevant.

The stellar mass from the derived SFH is $8.4\times10^{7}$M$_\odot$. Assuming a mass-to-light ratio of 1 or 2 to estimate the galaxy's stellar mass, together with its luminosity, we find values of $2.7\times10^{7}$M$_\odot$ and $5.3\times10^{7}$M$_\odot$, respectively. These masses are slightly lower than the first mass estimate: for this galaxy, we may be overestimating the low-mass end, given the SFR at ages older than $\sim4$ Gyr is particularly uncertain due to the low amount of information obtainable from the CMD.

\subsection{IC4247, ESO444-34} \label{sec_4247}

\begin{figure}
 \centering
  \includegraphics[scale=0.57]{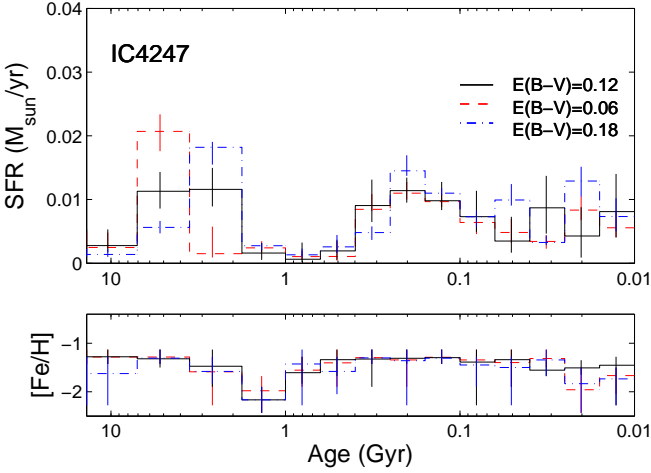}
 \caption{\footnotesize{Star formation rate and metallicity as a function of time for IC4247, as in Fig. \ref{sfhsnew}. This time, the SFH has been derived for three different input values of reddening in the code ($E(B-V)=0.06$, red dashed line, $E(B-V)=0.12$, solid black line, and $E(B-V)=0.18$, blue dot-dashed line).}}
 \label{sfh_redd}
\end{figure}

IC4247 has the highest stellar density in our sample, with a peak stellar density of 1040 stars per $0.1$ kpc$^{2}$. This galaxy is located at a deprojected distance of $\sim280\pm400$ kpc from M83, with a positive tidal index (see Tab \ref{infogen}). Its neutral gas content is $M_{HI}\sim3.5\times10^{7}$M$_\odot$, and this galaxy is detected also in H$\alpha$ \citep{lee07}. In the CMD in Fig. \ref{cmds}, the RGB, luminous AGB stars, blue and red helium-burning stars, and some upper MS stars are clearly visible, with a total number of $\sim18000$ recovered stars. As for ESO381-18 and ESO444-78, this galaxy has a high internal extinction due to its inclination ($A_B\sim0.74$ mag).

The recovered SFH is shown in Fig. \ref{sfhsnew}. The average SFR is $\sim0.0061\pm0.0045$M$_\odot$ yr$^{-1}$. This galaxy formed $\sim20\%\pm5\%$ of its stars prior to 8 Gyr ago and $\sim50\%\pm5\%$ of its stars before 5 Gyr ago. IC4247 had a rather constant SFH, with a period of slightly enhanced SFR at intermediate ages ($\sim2$ to 7 Gyr ago), and a period of very low star formation from $\sim400$ Myr to 2 Gyr ago. An estimate of the star formation within the last $\sim100$ Myr is given by the far ultraviolet (FUV) non-ionizing continuum. IC4247 is included in the sample of \citet{lee09}, who consider nearby luminous and dwarf galaxies to compare the SFRs derived from H$\alpha$ and FUV. In Fig. \ref{sfhsnew} we plot the SFR derived from FUV, and find it to be consistent with our results. At the present day, there is not very much star formation occuring as seen from the CMD, and we do not extend our SFH to ages younger than 10 Myr. The H$\alpha$ flux coming from the interstellar medium in the central region of the galaxy (taken from \citealt{lee07}) can be used to compute the current SFR. This value turns out to be $\sim0.0045$M$_\odot$ yr$^{-1}$, and we show it as reference in Fig. \ref{sfhsnew}, even though it refers to an age bin that we do not plot. We mention that, as for ESO381-18, it was rather difficult to match the color of the RGB with the other features in the CMD in the isochrone-fitting process. We thus had to use a value of $E(B-V)=0.12$ instead of 0.06, and the distance modulus was rearranged to $(m-M)_{0}=28.33$ instead of $28.48\pm0.21$, which is still within the errorbars. As discussed before, adjustments in the metallicity values would not lead to any major change in the derived SFH.

We decided to perform a test on IC4247 to estimate the influence of internal reddening on our SFHs. This galaxy has a high internal extinction due to its inclination, according to LEDA, although this value is just an estimate. We recompute the SFH for IC4247 changing the reddening value parameter in the code (see Sect. \ref{sfh_sec}). The reddening given by the \citet{schlegel98} maps is $E(B-V)=0.06$, so we perform the SFH recovery first adopting this value, and then using $E(B-V)=0.18$ as an extreme case of high reddening. By doing this, we moreover adjust the distance modulus value, in order to be able to reproduce the features of the observed CMD. The best-fit reddening value is $E(B-V)=0.12$ (as discussed above), but even substiantially changing this value does not affect dramatically our results. As shown in Fig. \ref{sfh_redd}, the results for the last $\sim1$ Gyr are similar within the errorbars, while the biggest differences are found at intermediate ages ($\sim2$ to 7 Gyr ago). This is because the RGB shape strongly depends on a combination of age, metallicity and reddening, and thus higher reddening values translate into younger ages, and vice versa. Overall, the best fit to the color of the MS is given by a value of $E(B-V)=0.12$, while the other reddening values give either too blue or too red sequences. The mean metallicity values are also consistent within the errorbars, although the highest reddening solution gives slightly lower mean values. We can thus safely conclude that the main results we obtain are not heavily affected by internal reddening.

The value of [Fe/H] looks constant over the history of IC4247. The mean value of [Fe/H]$=-1.37\pm0.21$ dex is slightly lower than the value we derive from the oxygen abundances of \cite{lee07} ([Fe/H]$=-1.03\pm0.20$), although still within the errors. The HII regions of this galaxy are rather small and concentrated in its central parts, and their enhanced metal content is the consequence of a recent short episode of star formation, which shows the inhomogeneity of the enrichment process. If the youngest populations have a higher metallicity, it is very difficult for us to recover this information just from the few MS stars, which are degenerate with age and metallicity and are blended with the BL at the faintest magnitudes of our CMD. More in general, even though the interstellar medium is enriched by star formation episodes, dwarf galaxies are not always able to retain the enriched gas. Furthermore, they are often not well mixed \citep[e.g., ][]{kniazev05, glatt08a, koch08, koch08b}. IC4247, with its high stellar density, possibly follows this trend, having some young ``pockets'' more enriched in the central regions where the potential is deeper.

From the SFH of IC4247 we estimate a stellar mass of $0.8\times10^{8}$M$_\odot$. Considering, instead, its luminosity and a stellar mass-to-light ratio of 1 or 2, we get values of $0.6\times10^{8}$M$_\odot$ and $1.3\times10^{8}$M$_\odot$, respectively, thus perfectly consistent with the previous estimate.

\subsection{ESO381-20} \label{}

\begin{figure}
 \centering
  \includegraphics[scale=0.23]{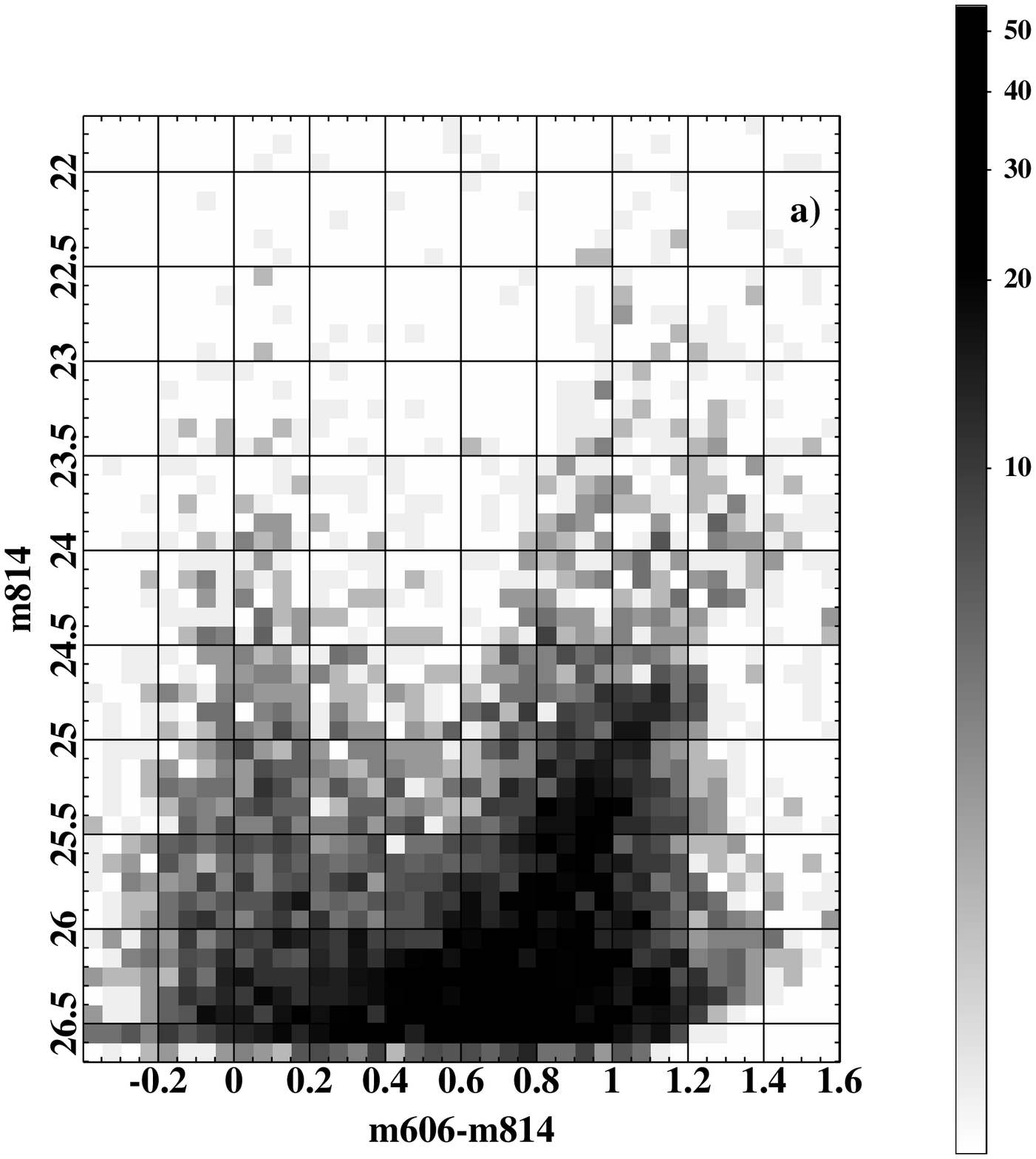}
  \includegraphics[scale=0.23]{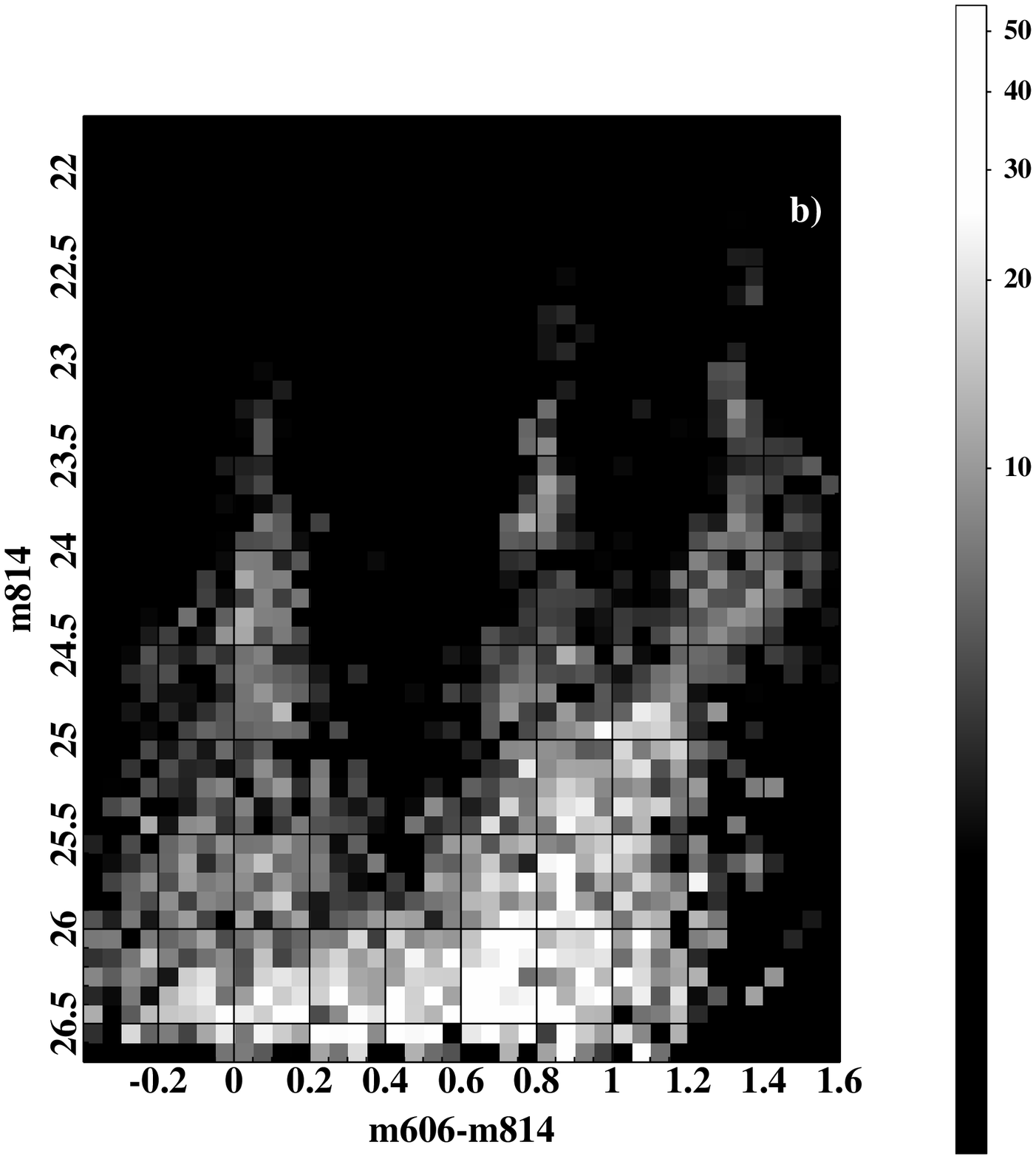}

\vspace{0.5cm}
  \includegraphics[scale=0.23]{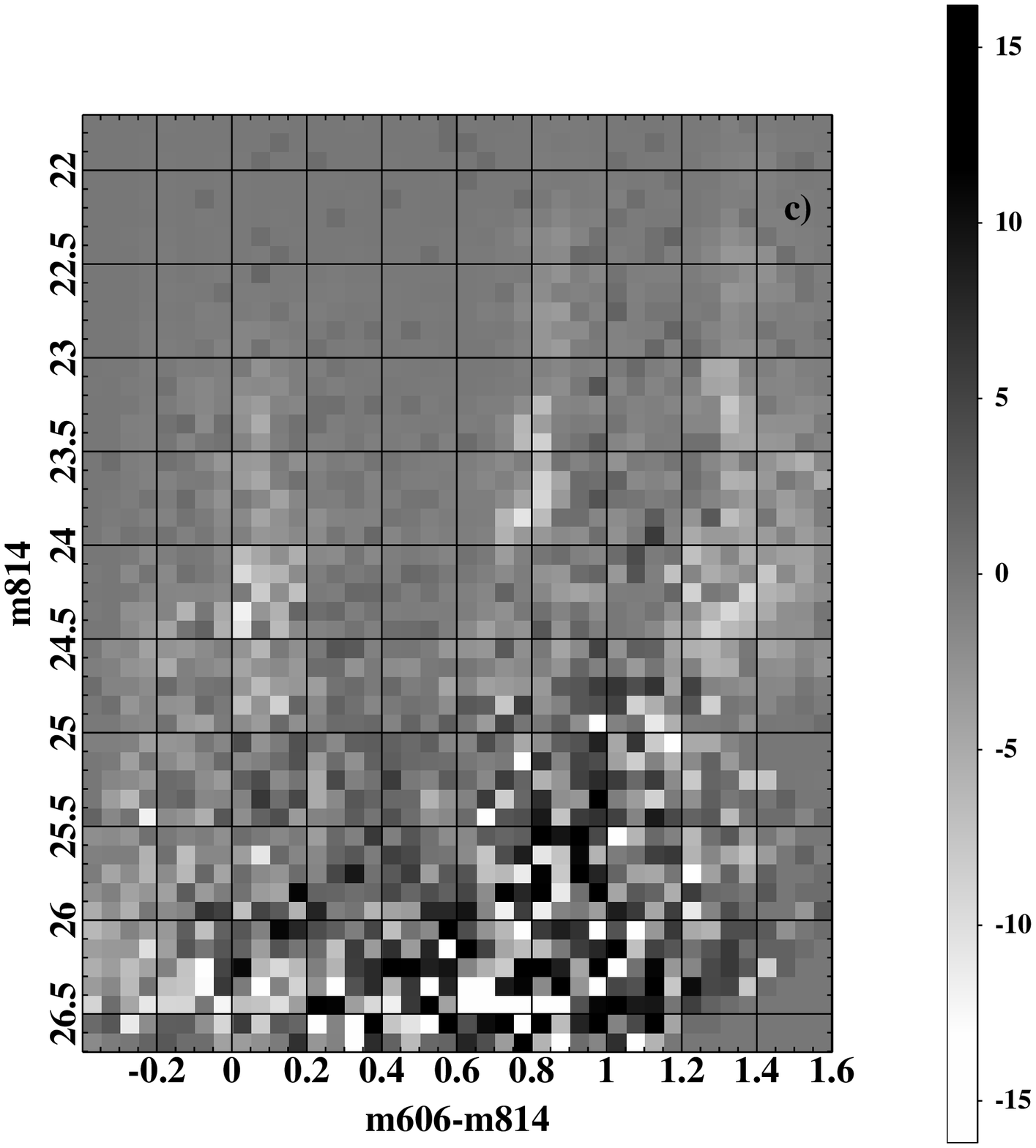}
  \includegraphics[scale=0.23]{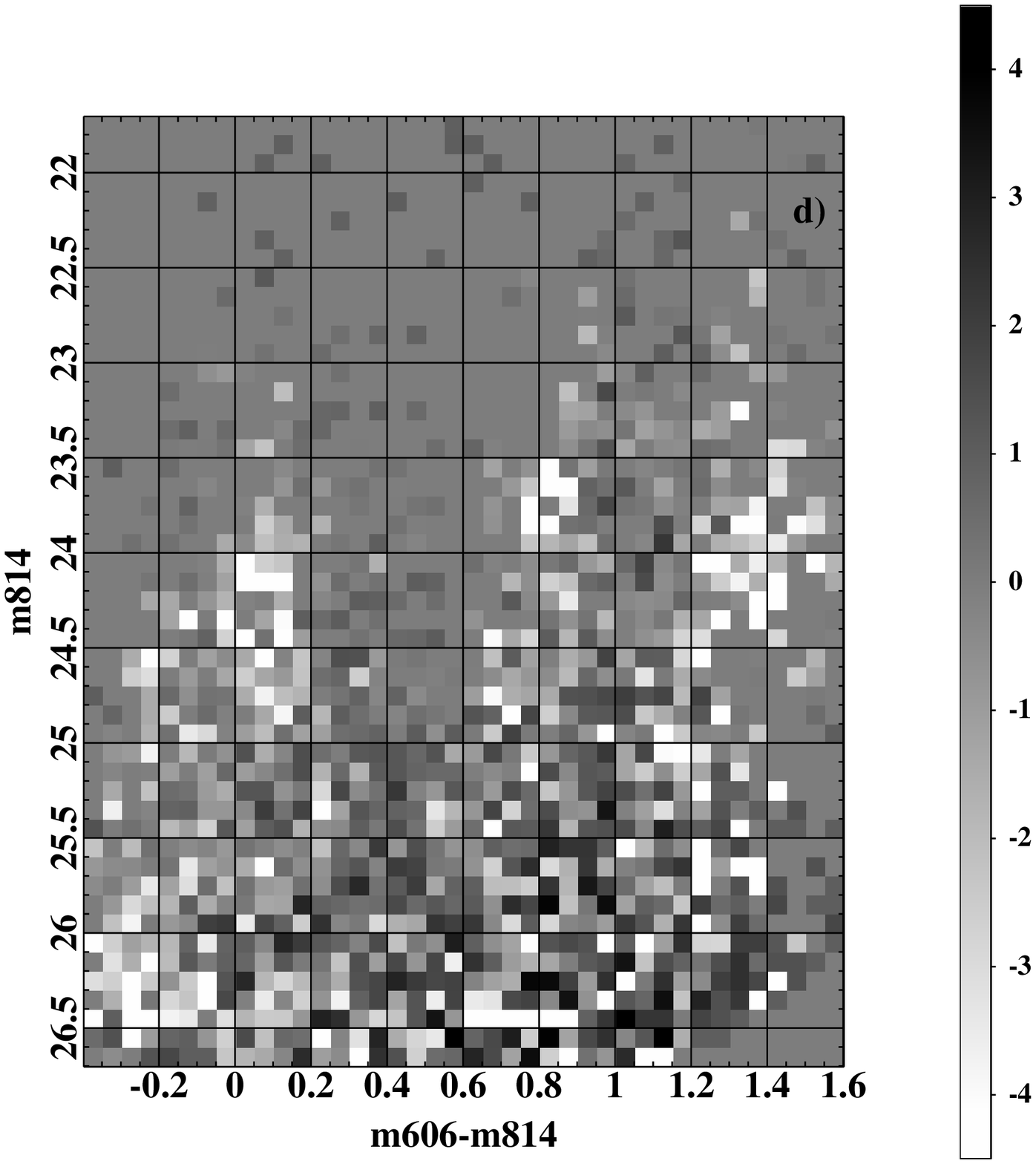}
 \caption{\footnotesize{Hess diagrams (displayed using ACS filters) for ESO318-20. The density values are listed along the colorbars. \emph{Panel a).} Data; the reported densities are in units of number of stars per bin (with a binsize of 0.05 mag in color times 0.10 mag in magnitude). \emph{Panel b).} Best-fit synthetic model, with a reverse color scale with respect to the data to facilitate comparison; units are number of stars per bin. \emph{Panel c).} Difference between data and best-fit model; units are number of stars per bin. \emph{Panel d).} Difference between data and best-fit model, weighted by the Poisson errors; units are proportional to the square root of the number of stars per bin.}}
 \label{38120pan} 
\end{figure}

The irregular galaxy ESO381-20 is located in the outskirts of the CenA/M83 group and is rather isolated (its tidal index $\Theta$ is negative). Its distance as found by \cite{kara07} is $5.44\pm0.37$ Mpc. Its deprojected distance from the closest massive neighbour (M83) is $\sim1.1\pm0.1$ Mpc. Its CMD (Fig. \ref{cmds}) contains $\sim20000$ stars and shows an old RGB, an intermediate-age luminous AGB and young MS, BL and RSG stars. This galaxy has a very high content of neutral gas ($M_{HI}\sim2\times10^{8}$M$_\odot$, see Tab. \ref{infogen}), which extends much further than the apparent optical galaxy. There is thus a high potential to form stars, and indeed its H$\alpha$ maps shows regions of very active star formation (\citealt{lee07}; \citealt{bouchard08}). Some of these star forming regions seem to coincide with local maxima in the HI distribution \citep{cote00}. The extinction due to the inclination is about 0.57 mag in the $B$-band (taken from LEDA). ESO381-20 is also the only galaxy within this sample that contains a globular cluster \citep{georgiev08}.

ESO381-20 seems to have experienced a relatively high continuous star formation during its lifetime (with an average of $\sim0.007\pm0.0048$M$_\odot$ yr$^{-1}$), which then increased substantially from 10 to 500 Myrs ago. This increased star formation activity was more than twice the average rate (see also the parameter $b_{500}$ in Tab. \ref{ressfh}). Following the recent discussion by \citet{mcquinn09}, we may definitely say that this is a period of a global starburst for the galaxy. The result does not change if we consider only the average SFR over the last $\sim4$ Gyr, as \citet{mcquinn09} do to avoid ``contamination'' from old ages. We will further discuss this starburst in the following Sect. We report in Fig. \ref{sfhsnew} the estimate of the SFR in the last $\sim100$ Myr, derived from the FUV continuum by \citet{lee09}, which is in good agreement with our results. For the most recent age bin (for this galaxy we are able to derive the SFR for ages as young as $\sim4$ Myr), the value derived here is consistent with the SFH derived by \cite{bouchard08} considering the H$\alpha$ flux of the galaxy ($\sim0.006\pm0.002$M$_\odot$ yr$^{-1}$, shown in Fig. \ref{sfhsnew}), but a factor of two higher than the H$\alpha$ SFR inferred by \citet{cote09}. After a visual examination of the HII regions selected in the two papers, we conclude that the discrepancy may be due to a slightly different selection of the emission regions, which are summed up to give the total flux. The star formation in the last few Myr has a value consistent with the average SFR from our SFH, as confirmed by the presence of only few MS stars compared to blue and red He-burning stars. We note that the SFRs derived from FUV and from H$\alpha$ differ from each other. \citet{lee09} show that, for dwarf galaxies, there is a systematic discrepancy between these two methods of deriving SFRs. In the case of ESO381-20, this seems to be due simply to the fact that the two tracers represent different star formation timescales. Already $\sim50\%\pm5\%$ of the stars of ESO381-20 were formed by 8 Gyr ago, while $70\%\pm5\%$ were in place 5 Gyr ago, but a substantial fraction was formed in the last Gyr (see Tab. \ref{ressfh}). 

Finally, in Fig. \ref{38120pan} we show the Hess diagrams for the observed CMD and the best-fit synthetic CMD. We can see that, given the higher number of stars for this galaxy, the resulting Hess diagrams are divided into smaller bins than in the case of ESO318-18 (Fig. \ref{38118pan}), from which the information needed to reconstruct the SFH is then evaluated. In the case of ESO381-20 the overall fit is better, with the biggest discrepancies between data and synthetic CMD being found in the less populated and most difficult to model regions (upper BL, upper RSG and luminous AGB).

Also in this case, the [Fe/H] value seems to be fairly constant during the entire galaxy lifetime. For comparison with our results, the oxygen abundances of the HII regions derived by \cite{lee07} are combined into a mean [Fe/H] value using the empirical formula by \cite{mateo98}, and the result is $-1.4\pm0.2$ dex, consistent with our value within the errorbars (see Tab. \ref{ressfh}). However, there are strong variations in the values for the individual HII regions (the nominal range is from [Fe/H]$=-1.87$ to $-0.90$). These findings would support the previous interpretation of the high metallicities found for the two HII regions of IC4247, in contrast with the value we derive from the SFH recovery.

For ESO381-20, a total stellar and dynamical mass were computed by \citet{cote00} via modeling of its rotation curve. They get a value of $2.3\times10^{8}$M$_\odot$ for the stellar mass, with a best-fit stellar mass-to-light ratio of 2.3 and a total mass-to-light ratio of about 20. From our method the result is $\sim1\times10^{8}$M$_\odot$, thus slightly lower. If we recompute the mass starting from the $B$-band luminosity and assuming a stellar mass-to-light ratio of 1 or 2, we get values of $\sim0.9$ and $\sim1.8\times10^{8}$M$_\odot$, respectively. This is consistent with our SFH. Since the depth of our CMDs is limited, it is plausible that in our resolved stellar study we are losing part of the mass coming from the oldest population, which results in a lower mass when compared to the intergrated light study of \citet{cote00}. As a check, we use the GALEV models \citep{kotulla09} with our derived SFH as an input to compute the resulting total luminosity of the galaxy, which turns out to be almost identical to the observed one reported in Tab. \ref{infogen}.

%________________________________________________________________

\section{Spatial distribution of stellar populations as a function of time} \label{maps_sec}

\begin{figure*}
 \centering
  \includegraphics[width=17.cm]{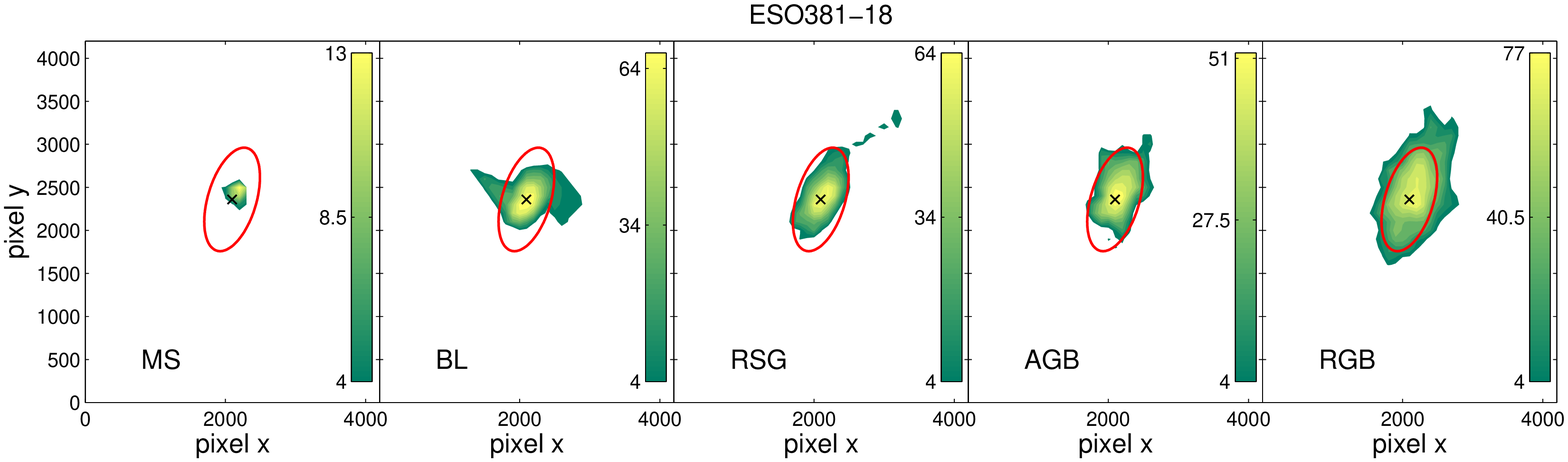}
  \includegraphics[width=17.cm]{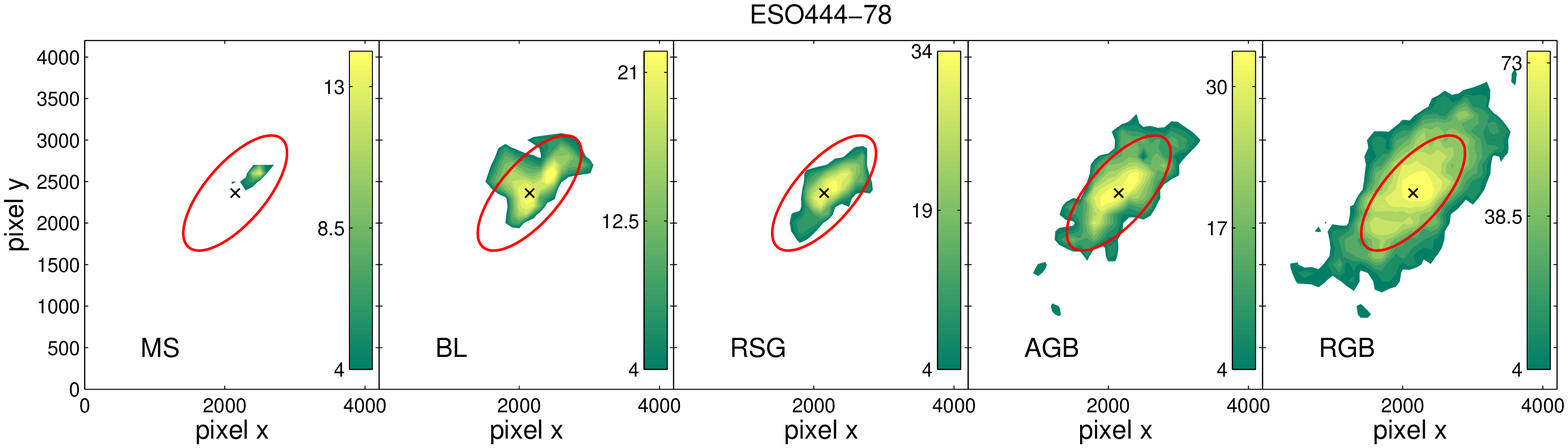}
  \includegraphics[width=17.cm]{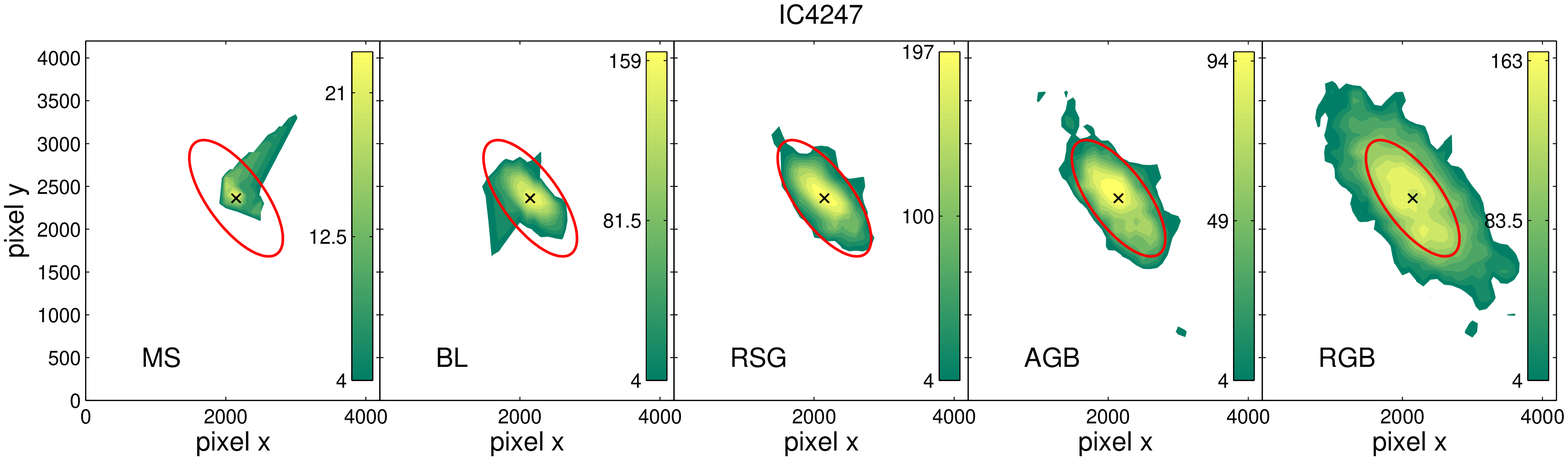}
  \includegraphics[width=17.cm]{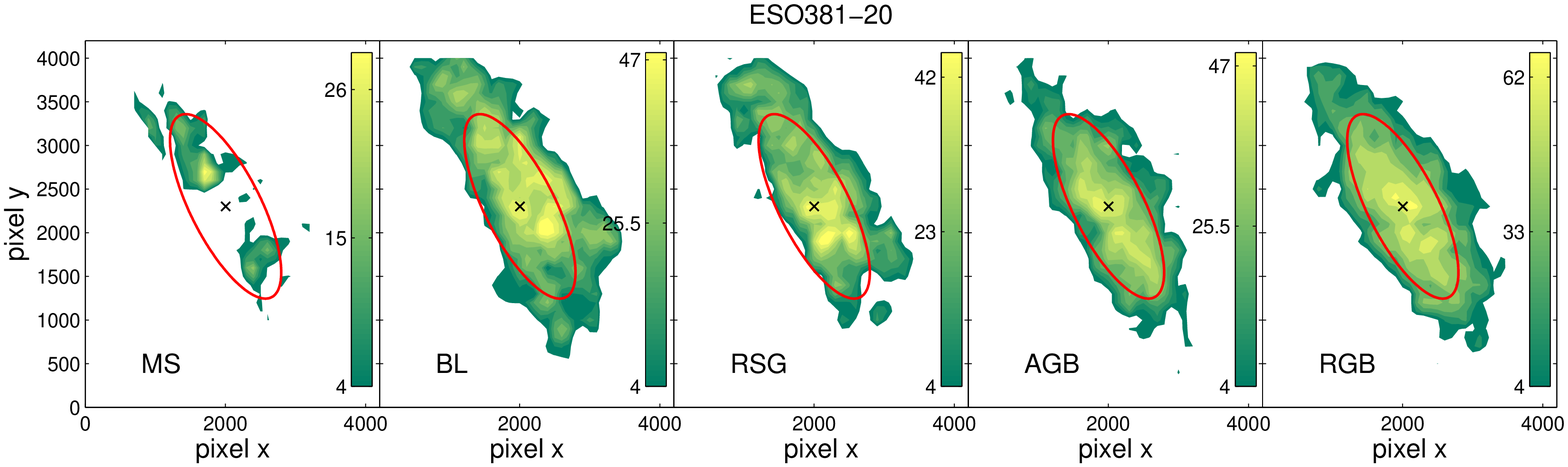}
 \caption{\footnotesize{Density maps for four of the M83 target galaxies (ESO381-018, ESO444-78, IC4247, and ESO381-20, ordered by absolute magnitude), each divided in five stellar evolutionary phases. These are: MS, BL, RSG, luminous AGB and RGB, ordered by increasing age. The stellar density values are listed along the colorbars, in units of number of stars per $0.1$ kpc$^{2}$. 10 equally spaced isodensity contours are drawn starting at the $1\sigma$ significance level up to the peak significance level. The peak levels are: for ESO381-018: MS=$2\sigma$, BL=$3.8\sigma$, RSG=$3.8\sigma$, AGB=$3.6\sigma$, RGB=$4\sigma$; for ESO444-78: MS=$2\sigma$, BL=$2.5\sigma$, RSG=$3\sigma$, AGB=$3\sigma$, RGB=$4\sigma$; for IC4247: MS=$2.6\sigma$, BL=$4.7\sigma$, RSG=$5\sigma$, AGB=$4.2\sigma$, RGB=$4.7\sigma$; for ESO381-20: MS=$3\sigma$, BL=$3.5\sigma$, RSG=$3.4\sigma$, AGB=$3.5\sigma$, RGB=$3.7\sigma$. The center of each galaxy is indicated with a black cross. Just as a reference among different frames, we also overplot in red the ellipse corresponding to the projected major axis radius at the isophote level 25 mag arcsec$^{-1}$ in $I$-band \citep[taken from ][]{sharina08}.}}
 \label{denmap2}
\end{figure*}

Dwarf irregulars are known to have scattered, clumpy regions of active star formation, with the less massive dwarfs only containing one such active region \citep[for a review, see ][and references therein]{grebel04a}. This is, for example, reflected in the shape these galaxies show when imaged in H$\alpha$. We now want to look for possible differences in the spatial distribution of stellar subpopulations. 

The subsamples into which the galaxies are divided were described in Sect. \ref{cmd_sec}, and are (ordered by increasing age): MS, BL, RSG, AGB and RGB. For each subsample, only stars with photometric errors smaller than $\sim0.1$ mag in magnitude and $\sim0.15$ mag in color are considered, since some features in the CMDs strongly overlap when the errors are larger (e.g., MS and BL stars, see Fig. \ref{isos}). This limit also corresponds to a $\sim80\%$ completeness level. We report in Fig. \ref{denmap2} the density maps of the five mentioned stellar subsamples for four of the five target galaxies. For ESO443-09 the number of stars is too small to draw significant density maps, so we do not report them.

\begin{figure*}
 \centering
  \includegraphics[width=10cm]{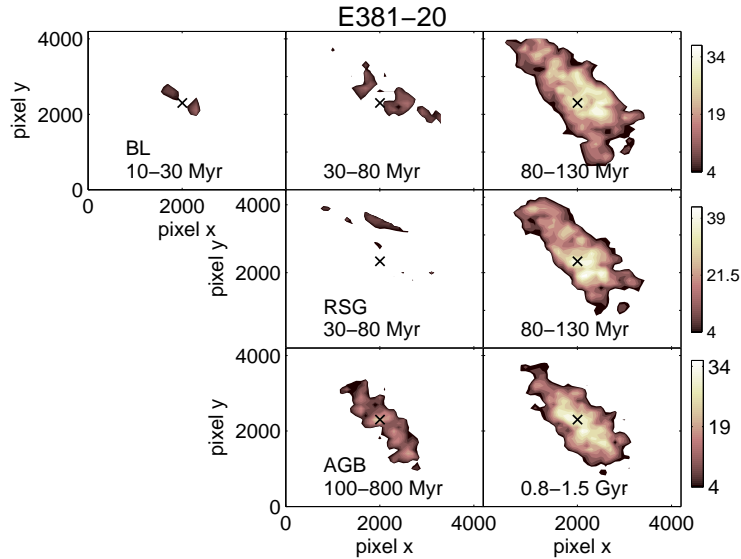}
 \caption{\footnotesize{Density maps for one of the target galaxies (ESO381-20), divided into different evolutionary stages. These are: in the top row BL stars; in the middle row RSG stars; in the bottom row luminous AGB stars. For each panel in each row, there is an age range as indicated. The color scale is the same within each row, normalized to the peak density of the densest (= oldest) subsample, and the stellar density values are listed along the colorbars (in units of number of stars per $0.1$ kpc$^{2}$). For each map there are 10 equally spaced isodensity contours, starting at the $1\sigma$ significance level up to the peak significance level. The peak levels are (from the youngest to the oldest sample): BL=$1.6\sigma,1.6\sigma,3.2\sigma$, RSG=$1.3\sigma,3.2\sigma$, AGB=$2.3\sigma,3.1\sigma$. The center of the galaxy is indicated with a black cross.
}}
 \label{spatsfh}
\end{figure*}

We compute density maps for the subsamples in the following way. The target galaxies are located at an approximate distance of $\sim5-6$ Mpc, at which 1 arcsec corresponds to $\sim0.03$ kpc. We assign to each star of a given evolutionary phase the number of neighbours found within $\sim0.07$ kpc$^{2}$ (a value chosen such that we do not add too much substructure but we still retain the overall features). We then convolve the result with a square grid. The final resolution of the density maps is of $0.02$ kpc$^{2}$. For each map there are 10 equally spaced isodensity contours. They start from a $1\sigma$ significance level, or $0.5\sigma$ where the number counts are too low (corresponding to $\sim4$ stars or $\sim2.4$ stars per $0.1$ kpc$^{2}$, respectively), and extend to the peak significance level, different for each map and indicated in the caption of Fig. \ref{denmap2}. The colorbars indicate the stellar density for each subsample, in units of stars per $0.1$ kpc$^{2}$. To facilitate comparisons among the stellar subsamples, we show with a black cross the center of the galaxy (i.e., coordinates listed in Tab. \ref{infogen}), and with an ellipse the projected major axis radius at the isophote level 25 mag arcsec$^{-1}$ in $I$-band \citep[taken from][]{sharina08}.

The youngest stars (MS in a range of $\sim10$ to 20 Myr and BL in a range of $\sim10-20$ to 150 Myr) are mostly concentrated in small ``pockets'' close to the the center of the galaxy. Given typical lifetimes of star forming complexes \citep[$\sim100$ Myr, see e.g.][and references therein]{dohm97}, it is reasonable to assume that these stars are still close to their birth place. They appear to form preferentially close to the central regions of the galaxies, where the potential is deeper. In some of the targets (e.g., ESO444-78 and ESO381-20) the most recent star formation episode took place in a region displaced from the center (as shown in Fig. \ref{denmap2}), while the BL stars reveal a similar off-centered activity region as well as activity in the galactic center. This resembles the distribution of recent star formation as seen in, e.g., Sextans A \citep{vandyk98}. For populations older than a few hundred Myr, the stellar distribution evens out. The RSG stars are approximately $\sim50$ to $\sim400-500$ Myr old and appear slightly more smoothly distributed. Finally, the intermediate-age and old populations (luminous AGB and RGB) are distributed basically over most of the galaxy's extent, with a quite regular elliptical shape. This reflects the long-time migration and redistribution of stars within the galaxy over long timescales. Such regular distributions of older populations appear to be a common trait of irregular galaxies \citep[e.g.,][]{zaritsky00, vandermarel01, battinelli07}. It is interesting to note that ESO381-20 has a very broad distribution for both BL and RSG stars. This is due to the fact that it experienced a strong burst of star formation in the last $\sim\!500$ Myr, that must have taken place across much of the spatial extent of the galaxy. We note that the RGB sample is likely to be contaminated by a fraction of intermediate-age and old AGB stars, which are less luminous than the TRGB.

We now discuss in more detail the helium-burning and luminous AGB samples. The position of a star in a CMD is determined, among other parameters, by age and metallicity, which in some evolutionary stages suffer a degeneracy, meaning that older and more metal-poor stars are found at the same location as younger and more metal-rich stars. This is particularly true in the RGB phase. Moreover, for the MS older and younger stars of the same mass occupy roughly the same position on the CMD. On the other hand, in their BL and RSG stages, at a fixed metallicity stars with different ages are well separated in the CMD (see Fig. \ref{isos}). Given that for our galaxies we are able to exclude a strong metallicity evolution with time (see previous Sect.), we can safely assume that the age-metallicity degeneracy is minimal in these stages and we can assign to each star in the BL and RSG phase a single age based on its position in the CMD. We want to use this information to better understand the distribution of stars at different lookback times, and see how the stellar populations are evolving.

We adopt the method described by \cite{dohm97} to separate the BL, RSG and AGB samples into older and younger subsamples, using Padova stellar isochrones as reference. We then again compute density maps as described before, this time for three age subsamples for the BL stars, and two subsamples for RSG and luminous AGB. This is done because it is easier to separate ages for the BL as compared to the other two evolutionary stages. We stress that for the luminous AGB phase the age determination for ages $>1$ Gyr is quite uncertain, since stars with different ages almost overlap in the region above the TRGB in $I$, $V-I$ color-magnitude space. We thus cannot evaluate precisely the age of the oldest AGB bin, and the age reported in the plot is just indicative. We report one example for this kind of density maps with high age resolution for ESO381-20, which is plotted in Fig.~\ref{spatsfh}.

For all of the studied objects, the youngest populations are again generally found in concentrated, actively star forming regions, while the stars with older ages are more broadly distributed. Theoretical expectations tell us that such complexes can have diameters up to several hundred pc in size, and contain OB associations and open clusters \citep[for a discussion, see Appendix A in][]{dohm97}. The former will quickly dissolve (within $\sim\!10-30$ Myr), but their stars will remain in the complex for $\sim\!50-100$ Myr, while open clusters are more long-lived and disrupt after several hundred Myr due to internal dynamics. It is not easy to put observational constraints on the lifetime of such complexes, because we are not able to follow them closely for long enough timescales. For Sextans A, \citet{dohm02} find a lower-limit age of $\sim\!100$ Myr, based on the spatially resolved SFH derived from BL stars. In our sample, ESO381-20 is the only galaxy that contains a substantial BL population, suitable for this kind of study, so we will now concentrate on this object.

\subsection{ESO381-20} \label{}

\begin{figure}
 \centering
%\resizebox{\hsize}{!}
  {\includegraphics[width=9cm]{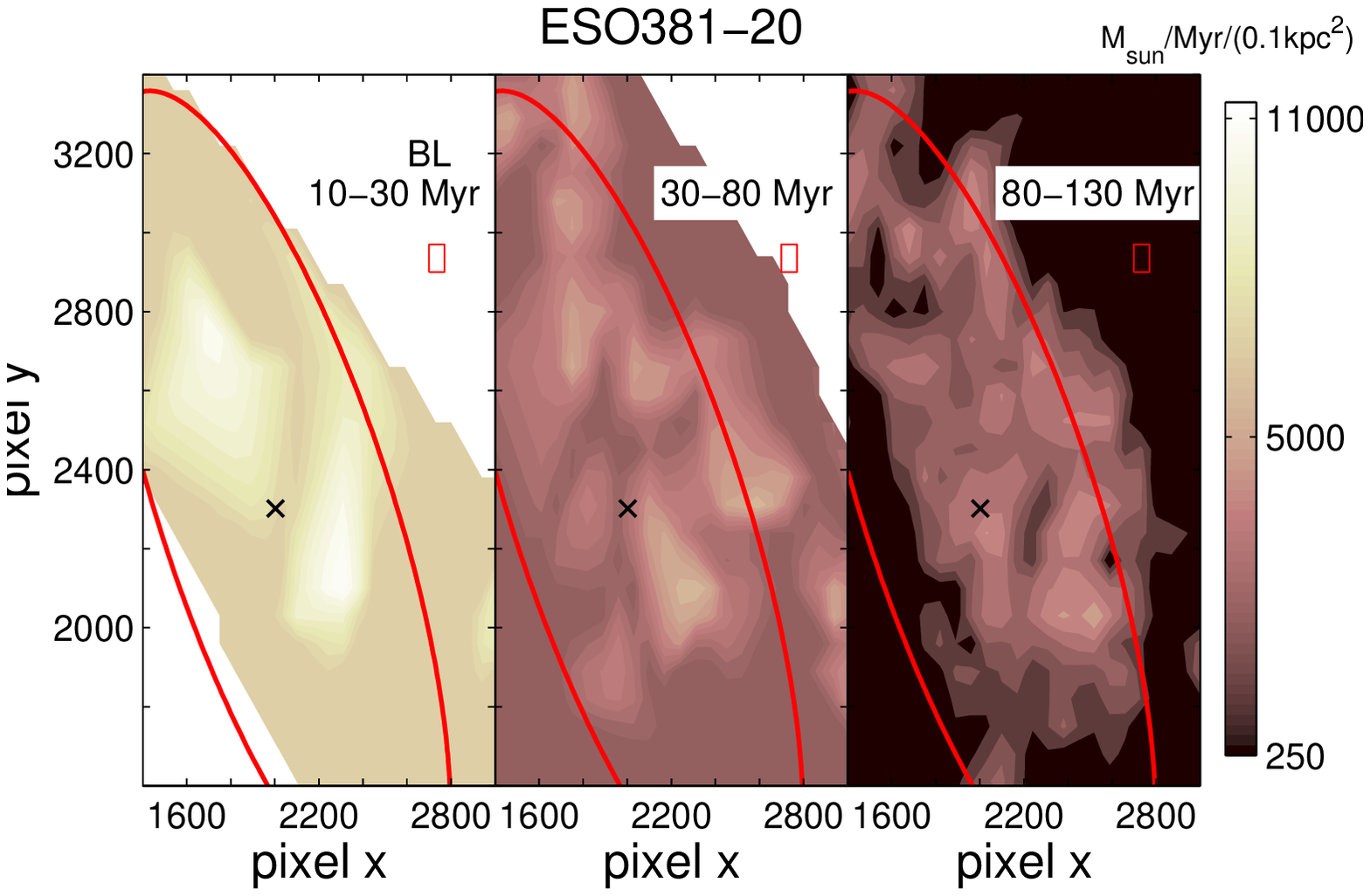}}
  {\includegraphics[width=9cm]{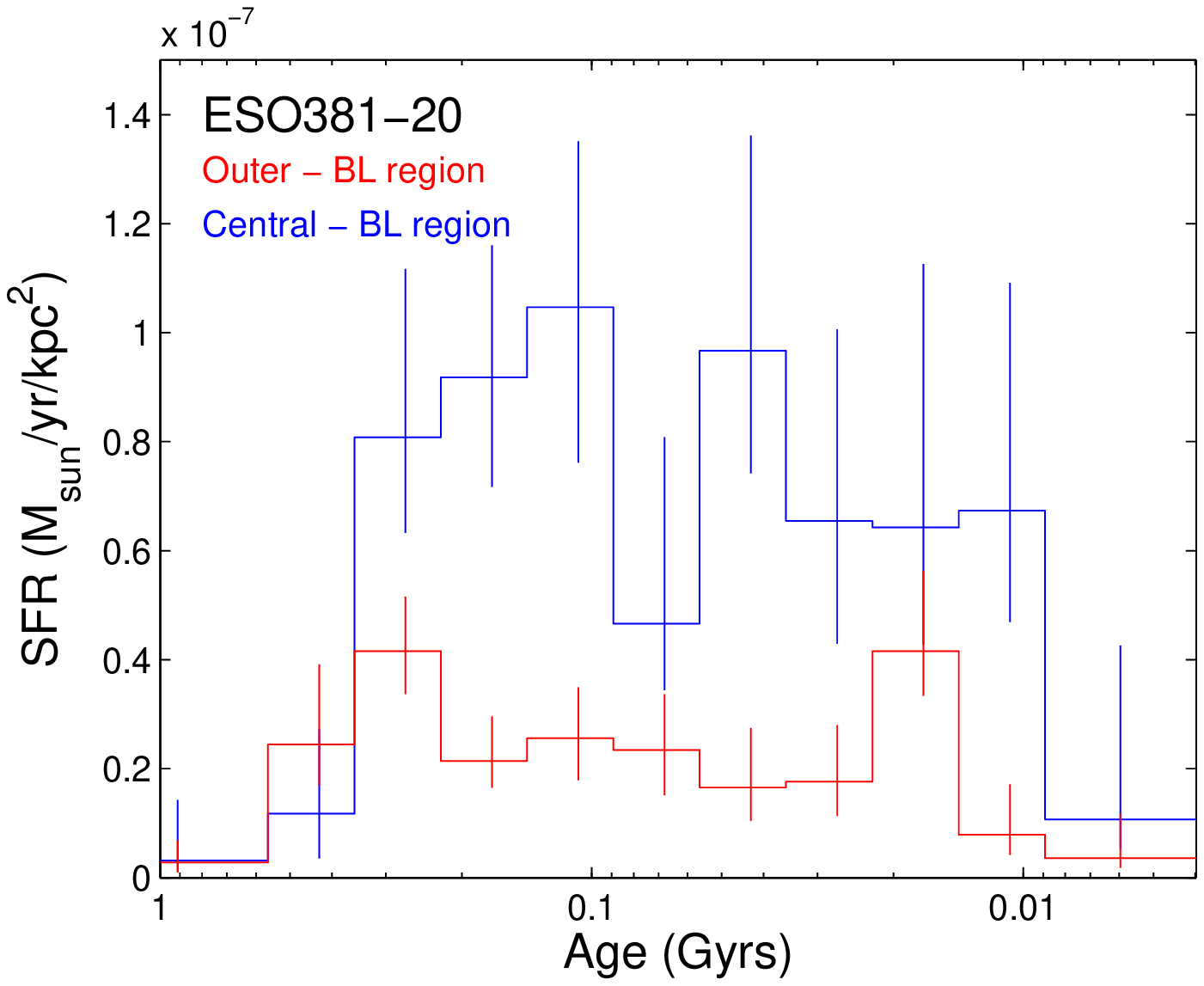}}
 \caption{\footnotesize{\emph{Upper panel.} Spatially resolved star formation history for the central region of ESO381-20, as from BL stars of different ages (as indicated in the subpanels). The SFR per unit area is indicated on the colorbar, with the same color scale for all of the subsamples. The angular resolution of the maps is marked with a red rectangle in each subpanel. The center of the galaxy is indicated by a black cross. Just as a reference among different frames, we also overplot in red the ellipse corresponding to the projected major axis radius at the isophote level 25 mag arcsec$^{-1}$ in the $I$-band \citep[taken from ][]{sharina08}. \emph{Lower panel.} Spatially resolved star formation history (via synthetic CMD modeling) of ESO381-20 within the last 1 Gyr. We select two subsamples of stars, found in the ``inner'' and in the ``outer'' regions of the galaxy, by looking at the density maps of the BL stars (Fig. \ref{spatsfh}, see text for details). The star formation rate for each region (normalized to the area of the considered region) as a function of time is plotted, with the oldest age being on the left side and the present day on the right edge of the (logarithmic) horizontal axis.}}
 \label{e381_sfh}
\end{figure}

ESO381-20 experienced a strong enhancement in star formation in the last $\sim500$ Myr, approximately three times higher than the average value. ESO381-20 also contains a higher number of BL stars as compared to the other galaxies in our sample, and thus we can take a closer look at its spatially resolved SFH. 

Fig. \ref{spatsfh} shows how the location of the peaks in stellar density in the BL phase changes with time. However, it is not straightforward to compare the density maps to each other. When sorting the subsamples by age, we are considering stars with different masses, and so the star formation needed to produce the observed densities will be different. We thus have to normalize the density maps for the IMF, and to do so we use the relations described by \citet{dohm97}. We zoom in on the central region of the galaxy, recompute the density maps for the BL subsamples, this time with a resolution of $0.01$ kpc$^{2}$, and finally normalize them to get units of M$_\odot$ Myr$^{-1}$ $0.1$ kpc$^{2}$. The results are shown in the upper panel of Fig. \ref{e381_sfh}.

We can see that the galaxy has kept forming stars throughout this age range, with several localized enhancements. This supports the idea of self-propagationg stochastic star formation \citep[see][]{seiden79, dohm02, weisz08}, for which the star formation is intermittently turning on and off in adjacent cells within the galaxy. It is not clear what the main mechanism is that triggers this mode of formation. Turbulence in the interstellar medium may locally enhance the gas density above the threshold for star formation \citep{elmegreen96}, or the reason could lie in the balancing process between heating (from stellar feedback) and cooling (inefficient at low metallicities) of the interstellar medium \citep{hirashita00}. These mechanisms would lead to large star forming complexes, with sizes of up to several hundred pc and lifetimes of $\sim100$ Myr. As mentioned before, some of the HII regions in ESO381-20 coincide with HI peaks. Looking at Fig. \ref{e381_sfh}, we can see how the peaks in star formation are moving as time proceeds, but unfortunately the low number of stars in the youngest bin only permits us to detect the two most prominent, intensively star forming regions. The peaks have diameters of $\sim100$ pc, thus consistent with the expectations. For reference, the physical distance from the center of the galaxy (black cross) to the highest density peak in the rightmost panel is $\sim0.8$ kpc. Their duration is more difficult to determine, given the small timescale resolvable. We can however notice that the peaks of star formation in the youngest age bin are already present, even though with a lower efficiency, in the middle panel, so activity there has lasted for at least $\sim80$ Myr. Several of the smaller peaks present in the oldest time bin are disappearing in the young ones, but adjacent cells are seemengly turning on. There does not seem to be any obvious spatially directed progression of the star formation, but from these data it is not possible to draw firm conclusions. A high-resolution HI mapping would be helpful to study more in detail the substructures in this galaxy \citep[see, e.g.,][]{weisz09}. We note that similar star formation characteristics and timescales have also been found in other irregulars, e.g., the LMC \citep{grebel98, glatt10}, and may be typical for these galaxies.

Moreover, star formation may be overall enhanced within galactic scales for long time periods ($\sim0.5-1$ Gyr, see e.g. \citealt{dohm97}, \citealt{mcquinn09}), due to a global starburst in the galaxy. ESO381-20 seems to have experienced such a high star formation period, and this was recent enough that we can try to look at the spatially resolved SFH within different parts of the galaxy, as derived from synthetic CMD modeling. Following \citet{mcquinn09}, we separate the galaxy in two smaller subregions, and we do that accordingly to the BL density maps in Fig. \ref{spatsfh}. We consider the inner region to be the one where the star formation activity has been high in the last $\sim100$ Myr, producing a high density of BL stars. The outer, currently less luminous, region is the one where stars younger than 130 Myr are found. We then want to see whether the enhancement in star formation has affected the galaxy as a whole, possibly with intermittent episodes, or whether it was localized in the central region. We compare the SFHs derived for these two subsamples, scaling them for the area of the regions, and plot them in the lower panel of Fig. \ref{e381_sfh}. The outer region is producing overall fewer stars with respect to the central one, but apart from that we do not see any obvious trend from the derived SFHs. On the contrary, the SFR is randomly varying with time around the average value. Even though the area normalized SFHs clearly show that the star formation in the faint region is on average more than two times lower than in the bright region, it still is enhanced by a factor of two with respect to the average SFR of the faint region itself. This simply tells us that the observed burst of star formation was a period in which the whole galaxy produced stars at a higher rate, and the phenomenon was not only localized to the currently bright central region.

%________________________________________________________________

\section{Discussion} \label{discuss}

\begin{figure}
 \centering
%\resizebox{\hsize}{!}
  {\includegraphics[width=9cm]{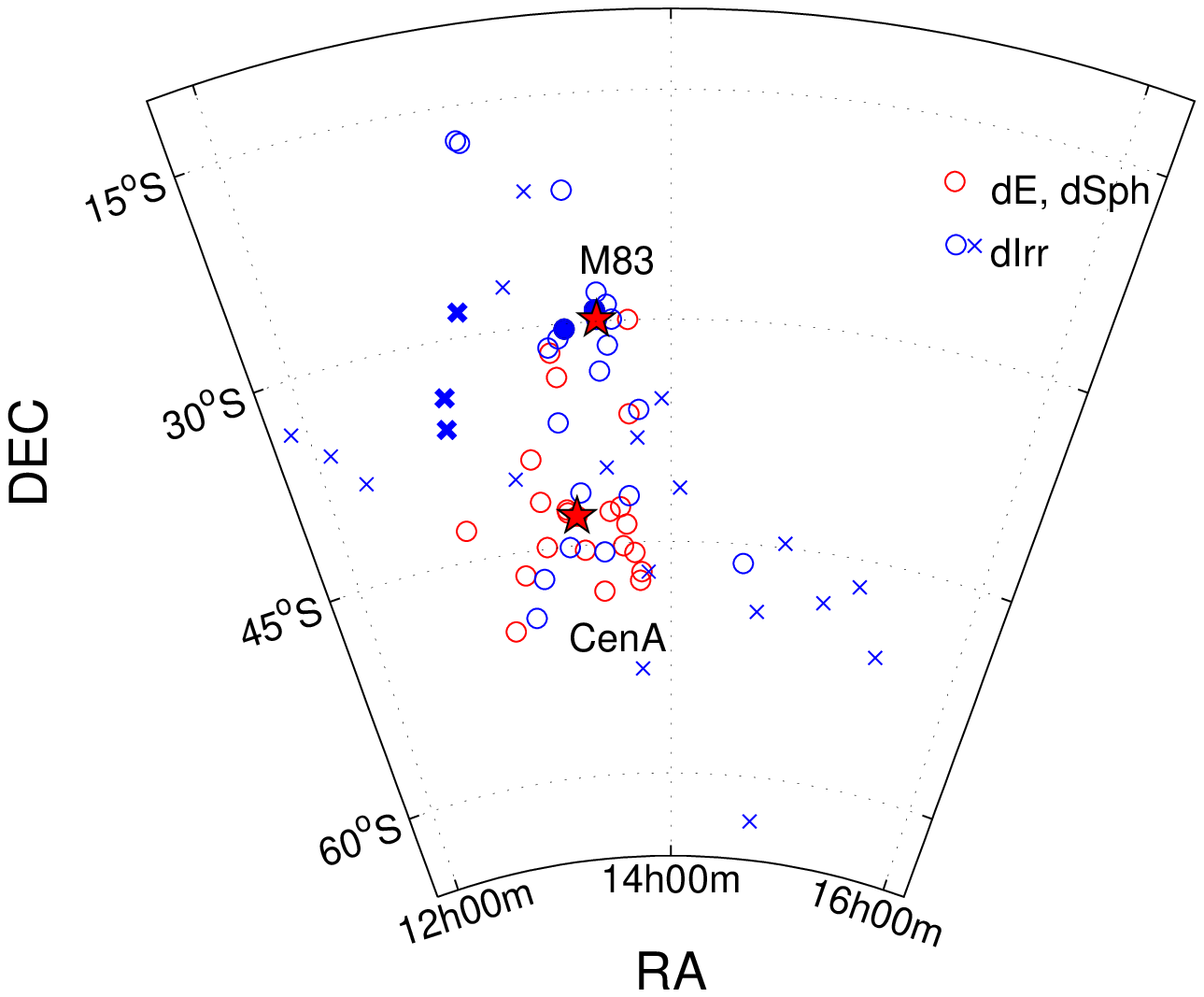}
  \includegraphics[width=9cm]{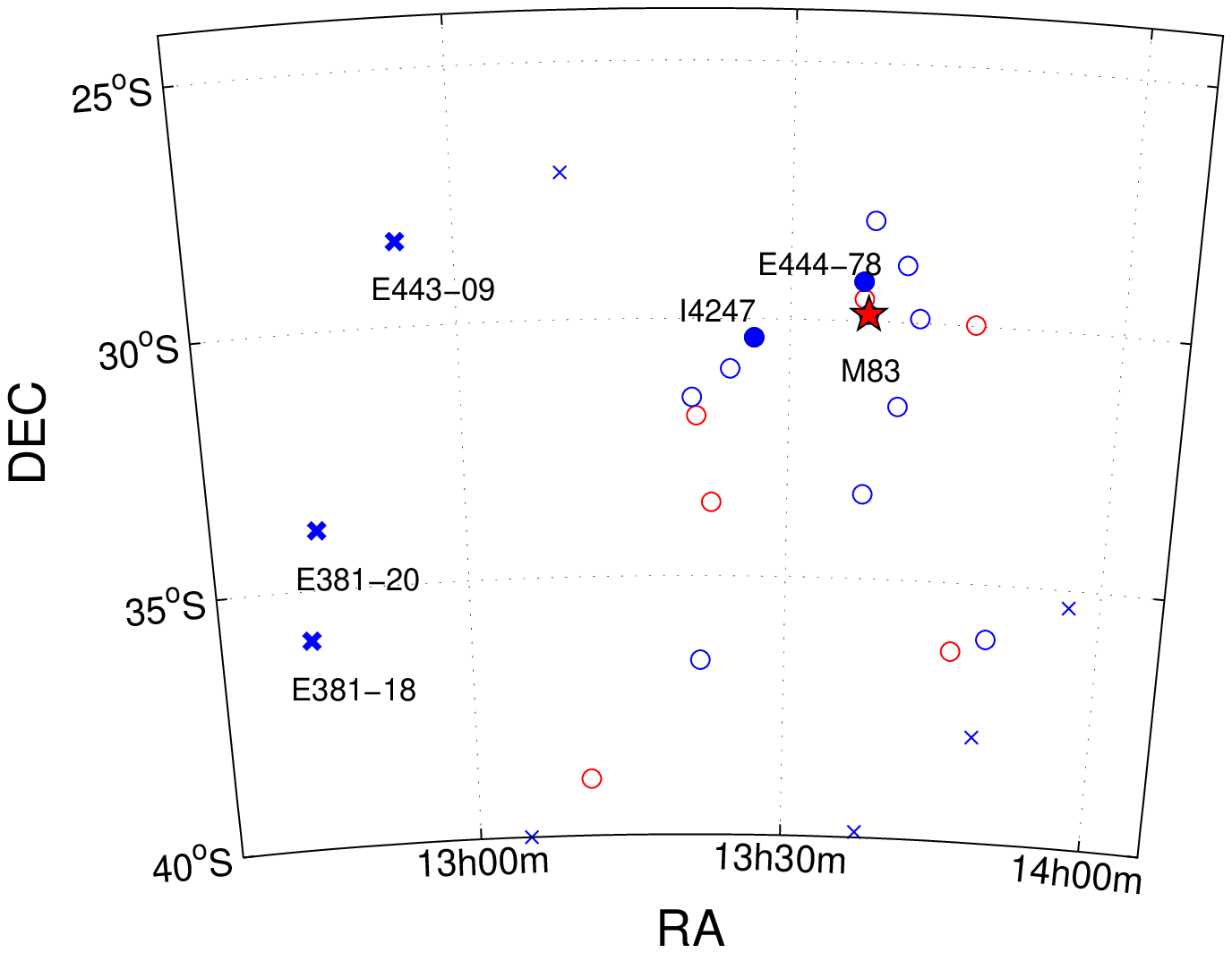}}
 \caption{\footnotesize{\emph{Upper panel.} Positions in the sky of the galaxies belonging to the Centaurus A/M83 complex (from \citealt{kara07}). Red symbols indicate early-type dwarfs (dwarf ellipticals, dE, and dwarf spheroidals, dSph), while blue symbols are for late-type dwarfs (dIrr). The circles are objects with positive tidal index, while the crosses stand for objects with negative tidal index. Two red stars are drawn at the positions of the two dominant giant galaxies CenA and M83, around which the smaller companions are clustering, forming two distinct subgroups. The filled circles and thick crosses represent the dwarfs studied here. \emph{Lower panel.} Same as above, just zoomed-in for a smaller region around M83, where the galaxies studied here are located (as labeled in the plot).}}
 \label{sky}
\end{figure}

Just as for the dwarf irregular galaxies of the Local Group, the target objects of our sample show considerable variety in their SFHs. The galaxies studied here cover a range of $\sim2.5$ mag, they have neutral gas masses of few $10^7$ to few $10^8$M$_\odot$, and the sample includes galaxies with both positive and negative tidal indices.

For almost all of our sample dwarfs, a period of old star formation ($\gtrsim5$ Gyr) at the lifetime average rate is followed by a lower-level activity for intermediate ages. We underline that episodes of enhanced star formation could be present also at these old ages, but the time resolution of our CMDs does not permit us to recover such episodes. Only in the last $\sim1$ Gyr we are able to resolve increased star formation activity above the average rate, for which ESO381-18 and ESO381-20 are the most striking examples. The enhancements in star formation found for all of the galaxies are usually a factor of 2-3 the average lifetime value. In contrast, IC4247 is an example of an overall rather constant SFH, with peaks of moderate intensity. In one single case (ESO444-78) the situation is very different: the star formation activity is high only for the first few Gyr, after which it stays rather constantly below the average value. \citet{cote09} suggest that this object could be a transition-type galaxy, given the absence of a strong H$\alpha$ emission. Its SFH resembles that of Local Group transition-type dwarfs like Phoenix or LGS3 \citep[see, e.g., ][]{dolphin05}, with two major differences. First, ESO444-78 is almost three magnitudes brighter than the mentioned LG dwarfs, and second, also its neutral gas content is higher by a factor of $10-100$ (see our Tab. \ref{infogen} and Tab. 1 in \citealt{grebel03}). However, we are not able to make a more detailed comparison between these objects, since our photometric depth does not permit us to resolve enhanced star formation episodes at ages older than $\sim5$ Gyr. 

The derived SFHs (Fig. \ref{sfhsnew}) seem to confirm the general trend found for objects in the Local Group, with quite long periods of star formation ($\sim100$ to 500 Myr) separated by quiescent epochs when the star formation is low but constantly active (``gasping'' regime, \citealt{marconi95}). The \emph{average} SFRs are of the order of $\sim10^{-3}$ to $\sim10^{-2}$M$_\odot$ yr$^{-1}$, which is slightly higher than the values found for Local Group dwarf irregulars in the same magnitude range, but comparable to the sample of objects in the M81 interacting group studied by \citet{weisz08}. Given the high activity seen for the giant galaxies in the Centaurus A group, \citet{cote09} also look for enhancements in the SFR of its dwarf members with respect to the Local Group, but they do not find evidence for this using the \emph{current} SFR. Finally, all of the galaxies in our sample seem to have formed at least $50\%$ of their stellar content before $z\sim1$ ($\lesssim8$ Gyr ago).

The position of our five galaxies in the Centaurus A group and a blow-up of the M83 subgroup are shown in Fig. \ref{sky} (the positions are taken from \citealt{kara07}). An important difference between the two subgroups is that the M83 subgroup, as opposed to the CenA subgroup, contains many more dwarf irregulars (shown in blue). Apart from the dwarfs that are likely bound members of the M83 subgroup, we also plot the positions of galaxies with negative tidal indices (i.e., isolated dwarfs), of which three are in our target sample. We want to compare the properties of the studied dwarfs considering also their position within the group. The currently most isolated objects are ESO443-09, ESO381-18 and ESO381-20, located at a deprojected distance of $>900$ kpc from M83. These galaxies appear to have undergone periods of enhanced star formation in the last few hundred Myr (relative to their lifetime average), thus showing that there is no need for a high density environment for these episodes to happen.

A puzzling property of the Centaurus A complex is the higher neutral gas mass to visual luminosity ratio of its members \citep{grossi07}, as compared to the Local Group or the Sculptor group, which are both less dense environments. When considering galaxy density, the stripping of the neutral gas from the dwarfs should thus in principle be more favored in the Centaurus A environment \citep[e.g., ][]{bouchard07}. This issue is extensively discussed by \citet{grossi07}, who are not able to find a definite answer. \citet{bouchard07} study the HI content of ESO444-78, and from an asymmetric elongation of the HI distribution they conclude that ram-pressure stripping could be occurring in for this object. When we compute the ratio of the present-day neutral gas content over the lifetime average SFR, we see that for both ESO444-78 and IC4247 (located at a deprojected distance of $\sim104$ kpc and $\sim280$ kpc from M83, respectively) it would take $\sim3.5$ and 6 Gyr to consume their entire HI amount at this rate, respectively. This could be a hint of a possible environmental effect on the neutral gas content of these dwarfs, since the galaxies of our sample that currently have a negative tidal index show instead high $M_{HI}/$SFR ratios ($\geq10^{10}$). Moreover, the $M_{HI}/L_B$ ratio for ESO444-78 and IC4247 is $\sim1.3$ and $\sim1.7$, respectively, while for the isolated dwarfs of our sample this ratio is always $\leq0.8$. \citet{lee07} and \citet{cote09} do not find any obvious trend of the neutral gas fraction as a function of tidal index or of distance from the dominant galaxy (the first work considers projected distances, while the second one uses deprojected distances), using data for many dwarfs in different nearby groups. On the other hand, \citet{bouchard08} find that both the neutral gas mass and the gas mass to visual luminosity ratio for dwarf galaxies are generally lower in denser environments, in agreement with our results, so the overall trend seems to be unclear.

In the previous Section, we computed parameters to quantify the amount of star formation for each galaxy in certain time periods, relative to the average value, and the fraction of stars born at recent, intermediate-age and old epochs (Tab. \ref{ressfh}). Our final aim is to investigate how these parameters behave as a function of luminosity, tidal index and deprojected distance from M83. Given the fact that the sample of galaxies considered in this paper is small, we postpone this discussion to a forthcoming paper (Crnojevi\'{c}, Grebel \& Cole 2011b, in prep.), where results for the entire sample of target galaxies will be presented.

Regarding the metallicity results, we emphasize again that with the current data we are not able to constrain the metallicity evolution with time and that our photometric metallicities are uncertain, but we can provide lifetime average values. All of the targets are metal-poor ([Fe/H]$\sim-1.4$ dex). This suggests that the galaxies may have been locally enriched within small-scale regions during periods of intense star formation, as can be seen from their H$\alpha$ emission, but they have afterwards experienced heavy loss of newly formed metals from star forming regions. The ejection can happen through galactic winds and SN explosions \citep[see e.g.][]{bradamante98}, and the dwarfs may have also accreted some primordial or little enriched gas during their lifetime. Alternatively, the enriched gas may still be in a hot phase, and thus not possible to detect at optical wavelengths \citep[e.g., ][]{recchi00, recchi06}. Previous literature studies also suggest that the metallicity-luminosity relation for external galaxies is similar to the one found in the Local Group. In particular, \citet{sharina08} consider the same dataset that we study, but as suggested in Sect. \ref{sfh_sec} their metallicities for the dwarf irregulars may be underestimated. We will thus also present a metallicity-luminosity relation for the entire sample of dwarf galaxies in the Centaurus A group in a forthcoming paper, including early-type dwarfs studied by \citet{crnojevic10} and the late-type dwarfs surrounding CenA.

%________________________________________________________________

\section{Conclusions} \label{conclus}

In the present work we analyze photometric archival HST/ACS data to study the stellar content of five late-type dwarfs in the Centaurus A group. The target objects show different luminosities, neutral gas contents and are located in different positions within the group (two close to the giant spiral M83 and three isolated objects in the outskirts of the group). We perform synthetic color-magnitude diagram modeling starting from Padova stellar isochrones \citep{marigo08}, and derive the star formation rate as a function of time for each object. The average star formation values range from $\sim10^{-3}$ to $\sim10^{-2}$M$_\odot$ yr$^{-1}$, which are typical values for low mass galaxies. The individual star formation histories appear to be very different from each other, but the overall trend is one where the star formation is taking place rather constantly over the galaxy's lifetime, with global enhancements of $2-3$ times the average value lasting up to $300-500$ Myr. Moreover, we find that for the target galaxies a fraction between $20\%$ and $70\%$ of the total stellar population was produced at ages older than $\sim7$ Gyr ago.

In a companion paper, we plan to investigate possible correlations between the parameters characterizing the star formation efficiency at different epochs and galaxy luminosity, degree of isolation, and deprojected distance from the dominant galaxy. For the current sample, we find that the ratio of neutral gas mass to average star formation rate is much lower for galaxies with positive tidal index than for those with negative tidal index, indicating that maybe the group environment has an effect on the neutral gas content of these dwarfs. These preliminary results will be completemented with a larger dataset.

Our results can be compared to late-type dwarfs in other group environments. The average star formation rates are similar to what is seen for dwarfs of comparable luminosities in the M81 group of galaxies, and slightly higher than those in our own Local Group. We find no significant differences among the three groups of galaxies when looking at the galaxies' individual star formation efficiencies at various time epochs, and also looking at the shape of the star formation histories themselves.

We present the stellar spatial distribution for stars of different ages within our galaxies. We confirm previous results about the oldest populations being more extended and having a regular shape with respect to the youngest stars, which show a more clumpy distribution. The latter is particularly pronounced for ESO381-20, which experienced a strong burst of star formation in the last 500 Myr. When looking at its resolved star formation history, traced by the youngest populations of core helium-burning stars, we clearly see that this prolonged period of high activity involved the galaxy as a whole, with several localized ($\sim100$ pc) and short-lived ($\sim100$ Myr) enhancements. We suggest that a stochastic mode of star formation takes place in these objects. The overall similar properties among the three galaxy groups, and the intrinsic scatter in the parameters characterizing the star formation, support this scenario. This will be further investigated in our forthcoming study of five additional late-type dwarfs in the Centaurus A group.

%________________________________________________________________

\begin{acknowledgements}

We thank the referee for his/her useful comments that helped to improve the paper. DC is grateful to S. Pasetto and S. Jin for helpful conversations and support, and thanks T. Lisker for his help with the GALEV models. DC acknowledges financial support from the MPIA of Heidelberg, as part of the IMPRS program, and travel support from the HGSFP of the University of Heidelberg. This work is based on observations made with the NASA/ESA Hubble Space Telescope, obtained from the data archive at the Space Telescope Science Institute. STScI is operated by the Association of Universities for Research in Astronomy, Inc. under NASA contract NAS 5-26555. This research made use of the NASA/IPAC Extragalactic Database (NED), which is operated by the Jet Propulsion Laboratory, California Institute of Technology, under contract with the National Aeronautics and Space Administration.

\end{acknowledgements}

%________________________________________________________________

\bibliographystyle{aa}
\bibliography{biblio.bib}

%________________________________________________________________

\end{document}